# Superconducting gap in covalent bismuth dihydride $BiH_2$ under extreme conditions

Jianning Guo[1,†], Dmitrii V. Semenok[2,*,†], Ivan A. Troyan[3,†], Pedro N. Ferreira[4], Di Zhou[2], Yulong Wang[1], Yuzhi Chen[1], Su Chen[1], Kexin Zhang[1], Xinyue Wu[1], Sven Luther[5], Toni Helm[5], Andrey V. Sadakov[6], Alexander T. Daniyarhodzhaev[6], Alexey S. Usoltsev[6], Leonid A. Morgun[6], Vladimir M. Pudalov[6], Christoph Heil[4], Viktor V. Struzhkin[3,7,*], and Xiaoli Huang[1,*]

[1] *State Key Laboratory of Superhard Materials, College of Physics, Jilin University, Changchun 130012, China*

[2] *Center for High Pressure Science and Technology Advanced Research (HPSTAR), Beijing, China*

[3] *Shanghai Key Laboratory of Material Frontiers Research in Extreme Environments (MFree), Shanghai Advanced Research in Physical Sciences (SHARPS), Pudong, Shanghai 201203, China*

[4] *Institute of Theoretical and Computational Physics, Graz University of Technology, NAWI Graz, 8010, Graz, Austria*

[5] *Hochfeld-Magnetlabor Dresden (HLD-EMFL) and Würzburg-Dresden Cluster of Excellence, Helmholtz-Zentrum Dresden-Rossendorf (HZDR), Dresden 01328, Germany*

[6] *V. L. Ginzburg Center for High-Temperature Superconductivity and Quantum Materials, P. N. Lebedev Physical Institute, Russian Academy of Sciences, Moscow 119991, Russia*

[7] *Center for High Pressure Science and Technology Advanced Research, Shanghai 201203, China*

* Corresponding authors, emails: dmitrii.semenok@hpstar.ac.cn (D. V. Semenok), viktor.struzhkin@hpstar.ac.cn (V. V. Struzhkin), huangxiaoli@jlu.edu.cn (X. Huang)

† These authors contributed equally.

## Abstract

Hydride superconductors at megabar pressures provide a promising platform for exploring room-temperature superconductivity. However, their superconducting gaps remain largely inaccessible to conventional spectroscopic due to diamond anvil cell constraints and minute sample dimensions. Here we develop a pulsed current method and apply it to covalent $BiH_2$ synthesized at 157-176 GPa. $BiH_2$ exhibits superconductivity at 58-70 K and upper critical fields of 11-17 T, substantially lower than those of many clathrate superhydrides, corresponding to a relatively long coherence length and an experimentally accessible critical current density. Short rectangular pulses minimize sustained Joule heating and enable currents up to 160 mA, allowing $J_c$(T) to be measured in the low temperature regime down to 2 K at 176 GPa. The normalized critical-current response remains reproducible between two measurement runs and is better described by a two-scale *s* wave model than by single-gap *s* or *d* wave models, yielding effective energy scales of approximately 6.9 and 1.5 meV. Fully anisotropic Migdal-Eliashberg calculations yield a single highly anisotropic gap, suggesting that the two-gap behavior observed experimentally originates from gap anisotropy rather than two independent gaps. These results establish pulsed critical-current measurements as a practical gap-sensitive transport probe under extreme pressure and, with further increases in peak-current capability, provide a route toward investigating room-temperature hydrides such as La-Sc-H.

## Introduction

Hydrogen-rich compounds have emerged as a major platform for high-temperature superconductivity under pressure. Following the discovery of superconductivity above 200 K in $H_3S$[1], a number of clathrate-like superhydrides, including $LaH_{10}$[2, 3], $YH_9$[4], $YH_6$[5], $CeH_9$[6], $CeH_{10}$[6], $ThH_{10}$[7], and $CaH_6$[8, 9], have exhibited exceptionally high transition temperatures at megabar pressures. In contrast to clathrate-like hydrides with atomic hydrogen frameworks, covalent and molecular hydrides such as $H_2S$[1], $SnH_4$[10], $SbH_4$[11], and $BiH_4$[12] possess distinct bonding environments and electronic structures, providing complementary platforms for investigating hydride superconductivity. Although their transition temperatures and upper critical fields can be characterized by electrical transport and magnetic field measurements, the microscopic gap information on their superconducting states remains scarce because the restricted geometry of diamond anvil cells (DACs) limits conventional spectroscopic and thermodynamic probes. At megabar pressures, gap measurements are further complicated by pressure gradients, phase coexistence, and the extremely small sample volume, which severely limit signal strength and spatial resolution. A major advance was recently achieved by Du et al., who fabricated planar $Ta/Ta_2O_5/H_3S$ tunnel junctions in DACs and directly resolved a fully superconducting gap of approximately 30 meV in $H_3S$[13]. This landmark work established tunnelling spectroscopy as a direct microscopic probe of superconducting gap at megabar pressures. However, implementing tunnelling spectroscopy in DACs requires specialized device fabrication, including the preparation of a well-defined tunnel barrier and stable junction conditions during high-pressure synthesis. Therefore, complementary gap-sensitive approaches compatible with conventional high-pressure transport configurations remain highly desirable.

The temperature dependence of the critical current, $J_c(T)$, provides an alternative route to superconducting gap information[14-19], which can distinguish between nodeless, nodal, and multiscale superconducting gap structures. For a superconductor in the self-field limit, $J_c(T)$ is governed by the magnetic penetration depth $\lambda(T)$, which is set by the temperature-dependent superfluid density and therefore by the superconducting gap $\Delta(T)$. Within the Talantsev-Tallon framework, the self-field critical current scales approximately as $J_c \propto \lambda^{-3}(T)$, so that the low-temperature form of $J_c(T)$ can be fitted using the corresponding BCS superfluid-density expressions to constrain the gap magnitude and symmetry[14, 15]. This analysis has been applied to the nickelate, and iron-based superconductors[16-19]. This makes critical current measurements a promising route for probing superconducting gaps under high pressure, where direct spectroscopic measurements remain difficult to implement. Unlike tunnelling spectroscopy, this approach does not require a deliberately fabricated tunnel barrier and can be performed using a conventional four-terminal DAC configuration. Measurements extending deep into the low-temperature regime are particularly informative because the low-temperature evolution of $J_c(T)$ provides enhanced sensitivity to the underlying gap symmetry including the fully gapped, nodal, and multiscale superconducting state. In practice, however, extending such measurements to low temperatures in high-temperature superconductors can face a practical obstacle: the critical current can become very large, making it difficult to probe directly. Driving such high currents through a DAC device generates considerable Joule heating, which may overheat and eventually destroy the thin electrodes and connecting leads. Consequently, extending $J_c(T)$ measurements deep into the low temperature regime requires a current delivery scheme that combines high peak current with a low time averaged thermal load. Short current pulses provide such a route by delivering the peak current required to reach the dissipative state while substantially reducing the duration and time-averaged power of the electrical excitation.

In this work, we synthesize covalent $BiH_2$ at 157-176 GPa and investigate its superconducting state through a combination of structural characterization, magnetotransport, and pulsed high-current measurements. The synthesized $BiH_2$ exhibits superconducting transitions at 61-70 K and upper critical fields of 11-17 T, markedly lower than those of many clathrate superhydrides and corresponding to a relatively long coherence length. These properties, together with the experimentally accessible current-induced suppression of superconductivity, establish $BiH_2$ as an ideal platform for demonstrating gap-sensitive critical-current measurements under megabar pressure. To overcome the severe Joule heating that hinders low-temperature transport studies in diamond anvil cells, we develop a short-pulse technique that reduces sustained heating and enables excitation currents up to 160 mA, thereby allowing the temperature dependence of the critical current density to be measured down to 2 K. Analysis of the resulting $J_c(T)$ curve supports a two-gap superconducting state, with gap values of approximately 6.9 and 1.5 meV. Fully anisotropic Migdal-Eliashberg calculations subsequently reveal a nodeless but strongly momentum-dependent gap, suggesting that the apparent two-gap response arises from pronounced gap anisotropy rather than two independent superconducting condensates.

## Results

### Structural characterization of analysis of the Bi-H systems

To investigate the superconductivity in Bi-H system, we loaded the elemental bismuth (Bi) along with ammonia borane ($NH_3BH_3$) in two diamond anvil cells (DACs B1, and B2) equipped with four electrodes in the van der Pauw scheme for electrical resistance measurements. A detailed description of the DACs design is given in Supporting Table S1.

The structure of the bismuth hydrides synthesized in DAC B1 was studied using X-ray diffraction at the Shanghai synchrotron research facility (beamline BL15, SSRF). The sample in DAC B1 was compressed to 157 GPa and laser-heated to over 1500 K, resulting in the synthesis of the new phase. The diffraction pattern (presented in Figure 1) has a pronounced point-like or "spotty" character, which corresponds to large microcrystals of bismuth hydrides, and contains at least two series of reflections. The first series (single diffraction line "Bi" in Figure 1b) corresponds to *bcc*-Bi (or Bi-V), which remains stable up to 300 GPa [20, 21]. The volume of the found *bcc*-Bi unit cell $V_{Bi}$ = 16.93 Å$^3$/Bi turns out to be 0.93 Å$^3$/Bi larger than the expected cell volume at 157 GPa (≈ 16 Å$^3$/Bi [21]). Hence, the most reliable pressure indicator in DAC B1 is the hydrogen vibron (4100 cm$^{-1}$, Supporting Figure S6), from which we determine the pressure to be approximately 140 GPa [22].

The second series of XRD reflections originate from $BiH_2$ (Figure 1a) whose heavy atom sublattice has a $Im\bar{3}m$ structure and whose hydrogen sublattice has a lower symmetry, probably $P2_1/m$, as theoretically predicted earlier[23]. The unit cell volume of the synthesized $BiH_2$ is $V_{exp}$ = 20.7 Å$^3$/Bi at 140 GPa, which is a bit larger (by 0.3 Å$^3$/Bi) than the theoretically predicted volume of $P2_1/m$-$BiH_2$ at 140 GPa ($V_{DFT}$ = 20.37 Å$^3$/Bi). X-ray diffraction of the DAC B2 sample at 176 GPa (Supporting Figure S9) showed the presence of small amounts of $C2/c$-$BiH_4$[12] and $P2_1$/m-$BiH_2$. Volumes of the synthesized phases are in good agreement with the published data[12, 24]: V($P2_1/m$-$BiH_2$) = 18.7 Å$^3$/Bi, and V($C2/c$-$BiH_4$) = 23.3 Å$^3$/Bi, confirming the reliability of the structural assignments.

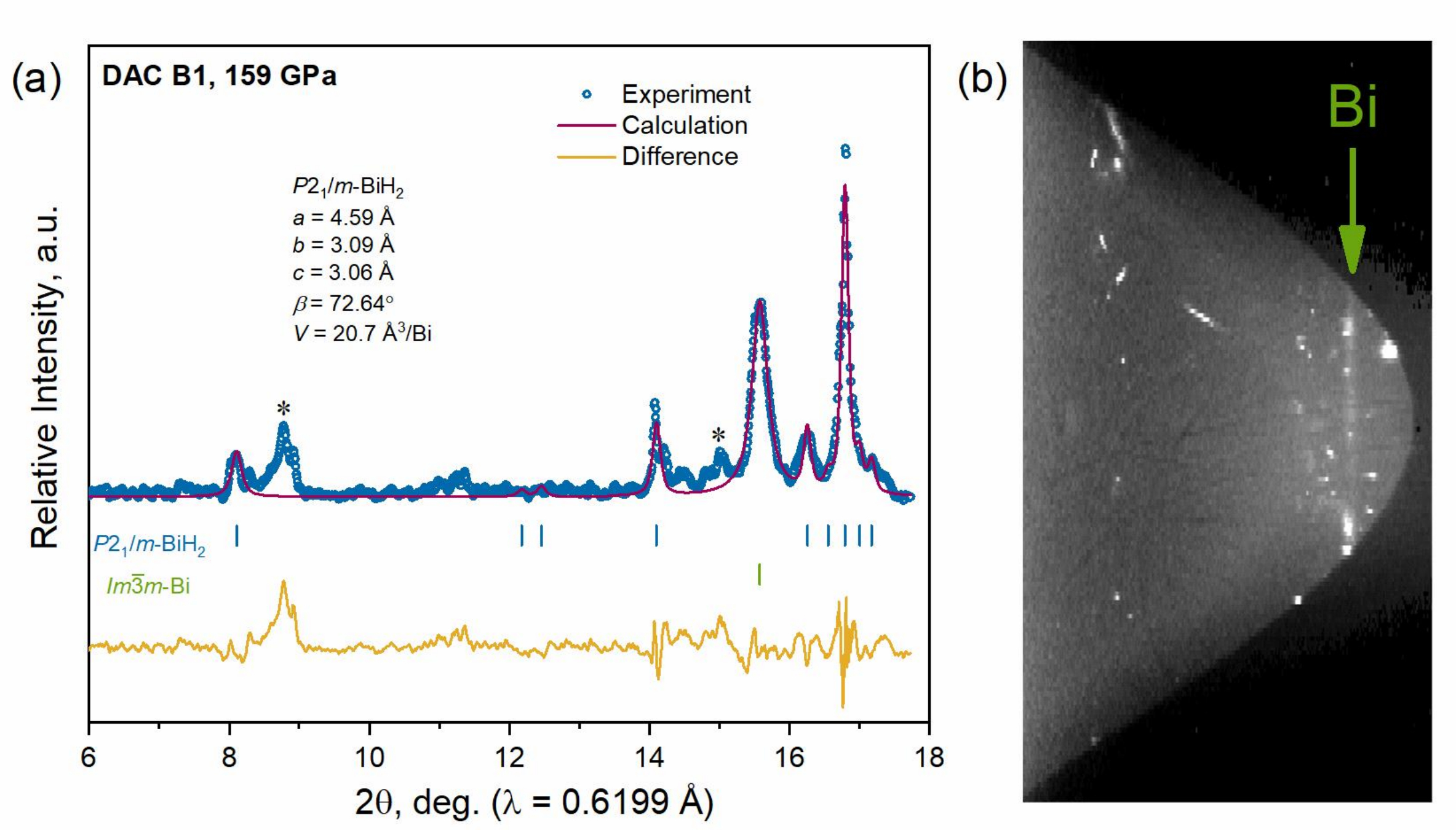

FIG. 1. X-ray diffraction patterns of the DAC B1 samples at 140 GPa (λ is 0.6199 Å) according to the hydrogen Raman shift and *bcc*-Bi volume. (a) Experimental pattern (blue circles), Le Bail refinement of the unit cell parameters (red line), and the residual signal (orange line). (b) Typical spot-like diffraction pattern of the sample in DAC B1, indicating a rather large (~1 μm) crystallite size of $BiH_2$. There is also a significant amount of unreacted Bi (marked "Bi").

## Superconductivity and transport properties of $BiH_2$ and $BiH_4$

The sample configuration and electrode geometry for each diamond anvil cell (DAC) are illustrated in Figure 1a. The electrical transport measurements in independently DACs reveal a reproducible superconducting transition between approximately 59 and 70 K, which we attribute to $P2_1/m$-$BiH_2$. In DAC B1, partial resistance drops were observed at 65-70 K in the pressure range of 159-163 GPa, with the highest transition temperature of approximately 70 K at 159 GPa (Figure 1b). Repeated laser heating slightly modified the transition profile but did not eliminate the finite residual resistance. The residual resistance below 18 K is most likely attributable to unreacted elemental Bi[25, 26], consistent with the X-ray diffraction results.

To achieve a zero-resistance state associated with $BiH_2$, the sample in DAC B2 was subjected to multiple laser-heating cycles. At 169 GPa, the resistance initially began to decrease at approximately 90 K, followed by a distinct shoulder near 61 K (Figure S1c). Across the transition region, the resistance decreased by nearly three orders of magnitude, although a finite residual resistance and a small anomaly at 16-18 K remained at low temperature. The latter is most likely associated with unreacted elemental Bi. The coexistence of several resistance features is consistent with the multiphase character of the sample revealed by X-ray diffraction.

Over several months of subsequent transport measurements, the resistance profile changed substantially. The high-temperature onset near 90 K gradually weakened and eventually became barely detectable, whereas the transition near 61 K became sharper and accounted for an increasingly large fraction of the total resistance drop (Figure 2c). After a slight pressure increase to 176 GPa, the low-temperature resistance associated with this transition fell below the experimental noise level, corresponding to an essentially zero-resistance state (Figure 2c). Similar behavior was detected using two perpendicular voltage-contact configurations, indicating that the pronounced resistance decrease was not restricted to a single local current path. The persistence and progressive enhancement of the 58 K transition, despite the disappearance of the higher temperature feature, provide further evidence that this superconducting channel originates predominantly from $BiH_2$. The evolution of the resistance curves likely reflects a gradual redistribution of the conducting phases and current pathways within the sample chamber.

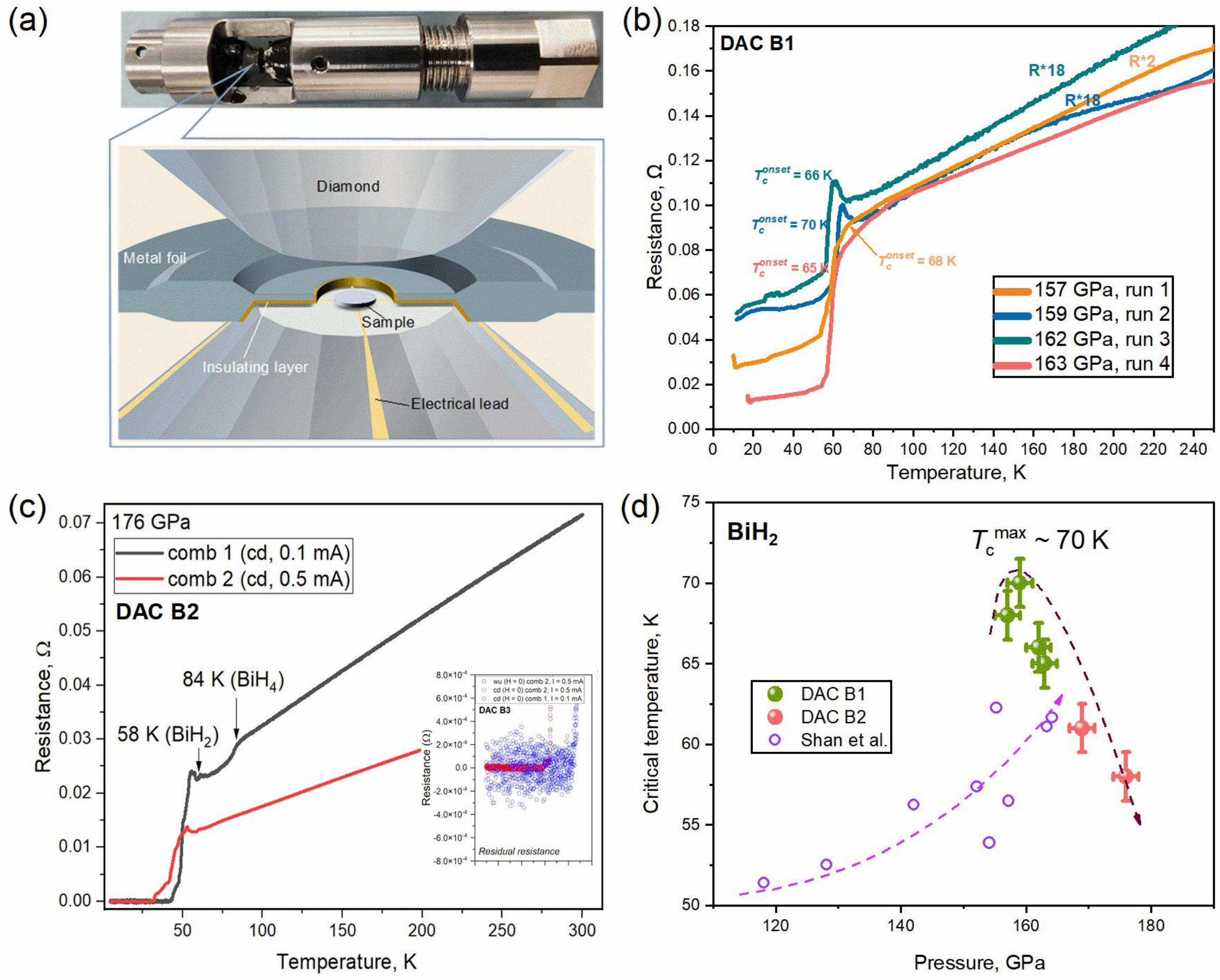

FIG. 2. Electrical transport study of $BiH_x$ samples at different pressures. (a) photo of the DAC B1 and sketch of the sample's positioning on the electrodes in the high-pressure chamber of DAC B1. (b) Temperature dependencies of the electrical resistance in DAC B1 after a series of laser heating steps (runs 1-4). Resistance was scaled (R*2, R*18) for ease of comparison. (c) Temperature dependence of the electrical resistance of DAC B2 sample after degradation and possible decomposition of $BiH_4$ (176 GPa). "Comb 1,2" mark perpendicular combinations of electrodes used for the voltage measurements in the van der Pauw scheme. (d) The $T_c$'s dependence on pressure for $BiH_2$ superconducting phases. The purple circles represent the results obtained by Shan et al. for $BiH_2$[24].

## The upper critical magnetic fields of $BiH_2$

To determine the upper critical field, electrical transport measurements were performed on DACs B1 and B2 in steady magnetic fields. In both cells, the resistance transitions shifted systematically to lower temperatures with increasing fields, confirming their superconducting origin. Throughout this work, $T_c$ was determined using the onset criterion. The resulting $H_{c2}$(T) curves exhibit near-linear with a slight upward curvature, a feature commonly observed in multiband superconductors. We therefore fitted the data using a two-band extension of the Werthamer-Helfand-Hohenberg (WHH) model[27]. For $BiH_2$ in DAC B1, the extrapolated upper critical magnetic field is $\mu_0H_{c2}(T) = 16$ T. And for DAC B2, the fitted upper magnetic fields are 14 T (169 GPa) and 11.5 T (176 GPa), respectively. These values are substantially lower than the corresponding Pauli paramagnetic limits of approximately 119 T, suggesting that the upper critical field is primarily governed by orbital pair breaking. Using the Ginzburg-Landau relation $\mu_0H_{c2}(T) = \Phi_0/2\pi\xi^2(0)$, the corresponding coherence lengths are estimated to be above 45 Å. Thus, despite its relatively high superconducting transition temperature, $BiH_2$ possesses a comparatively low upper critical field and a relatively long coherence length.

The upper critical field was further examined using pulsed magnetic fields up to 50 T in DAC B1 (Figure S3). At 2 K, the superconducting transition was suppressed at approximately 8.8-10.5 T, in reasonable agreement with the extrapolation from steady field measurements. The slightly smaller value obtained in pulsed fields may be associated with transient heating and uncertainties caused by the rapid field sweep. The agreement between steady and pulsed field measurements confirms the comparatively low upper critical field of $BiH_2$. Together with its experimentally accessible transport critical current, this relatively low field scale makes $BiH_2$ a suitable platform for the pulsed high current measurements discussed below.

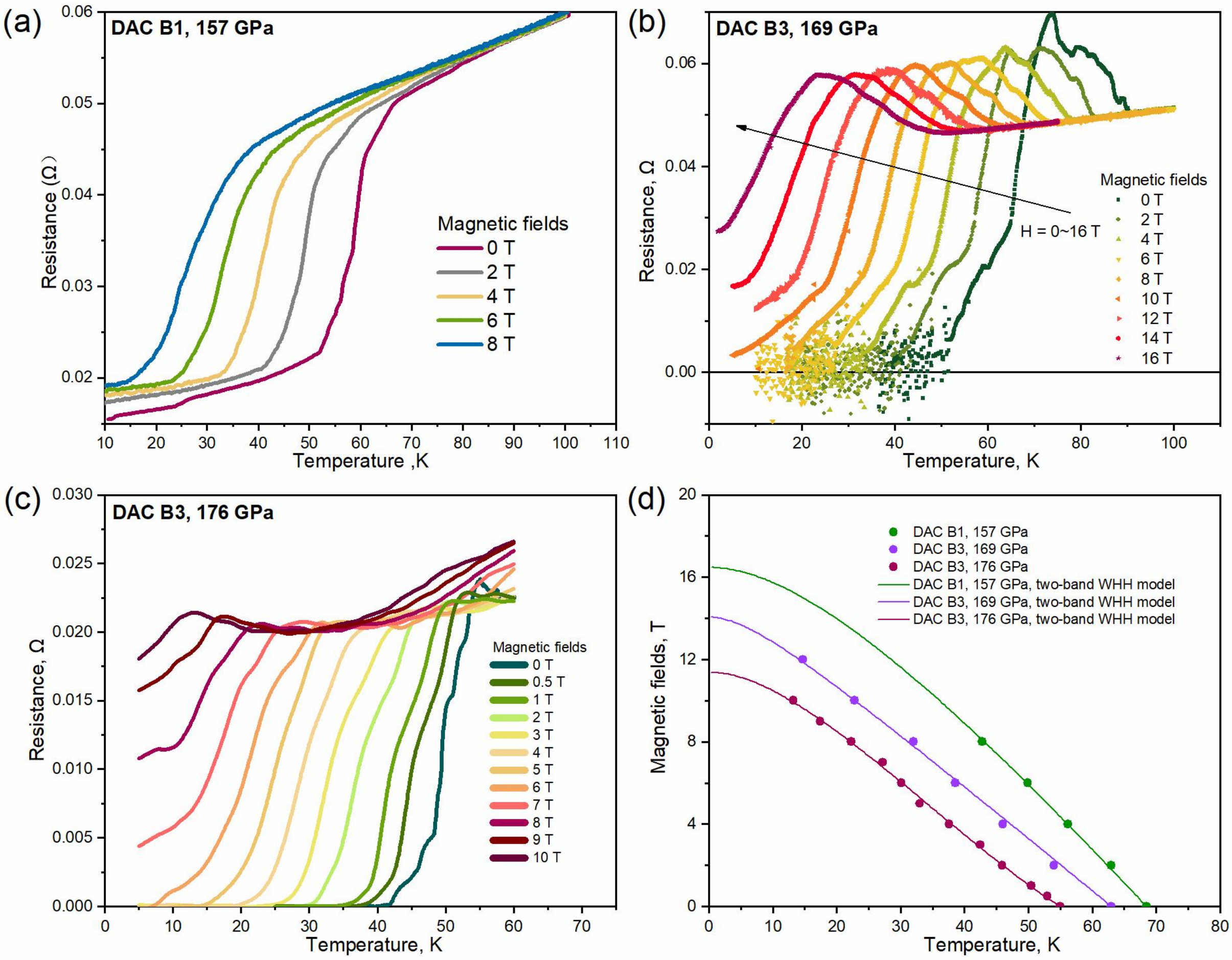

FIG. 3. Experiments with bismuth hydrides in steady magnetic fields. (a), (b) and (c) The temperature dependence of the electrical resistance of $BiH_2$ (DAC B1 and B2) under applied magnetic fields. (d) Critical magnetic fields obtained by fitting the experimental data with the two-band WHH model under steady-state magnetic fields. For all DACs, the $T_c$(onset) was used to draw this diagram.

Comparing the superconducting transition temperatures and upper critical fields of various binary hydrides (Supporting Figure S4c), we find that many covalent hydrides exhibit relatively small $\mu_0H_{c2}(0)/T_c$ ratios, typically below 0.5 T/K. Covalent hydrides generally show lower upper critical fields than clathrate-like hydrides with comparable $T_c$. For example, $H_3S$[1] and $SbH_4$[10] exhibit transition temperatures similar to those of $C2/m$-$LaH_{10-x}$[28] and $Fm\overline{3}m$-$CeH_{10}$[6], but their $\mu_0H_{c2}(0)$ values are approximately two to three times smaller. For $BiH_2$, $T_c = 55$ K and $\mu_0H_{c2}(0) = 11.5$ T at 176 GPa, corresponding to $\mu_0H_{c2}(0)/T_c = 0.2$ T/K. This value is substantially lower than those of cubic $LaH_{10}$, with $T_c = 250$ K and $\mu_0H_{c2}(0)/T_c = 0.5$ T/K[2], and $CaH_6$, with $T_c$ =210 K and $\mu_0H_{c2}(0)/T_c = 1.2$ T/K[8, 9]. Moreover, clathrate-like $CeH_9$[6], $Lu_4H_{23}$[29], and $La_4H_{23}$[30], despite having transition temperatures comparable to that of $BiH_2$, exhibit upper critical fields above 30 T. These comparisons highlight the comparatively low superconducting field scale and relatively long coherence length of $BiH_2$.

We can understand the surprisingly low value of $\mu_0 H_{c2}(0)$ in hydrides of $p$-elements with the high Fermi velocity. The Bi atom has an odd number of electrons in the outer shell ($6p^3$). The same is true for Sb ($5p^3$) and P ($3p^3$), all three elements have their outer electron shell filled exactly by half. With such half-filling, both the orbital moment and the total moment J = L + S = 3 reach their maximum values. In contrast, the outer shell of Y and La ($s^2d^1$) is filled by only 10%, and the total moment $J$ is close to its minimum value. For orbitals with $p$-symmetry, there is a large contribution of the spin-orbit coupling (SOC) to the thermodynamic and structural properties of compounds. Moreover, this contribution is absent for the symmetric $s$-shell. The contribution of SOC in bismuth compounds is especially large due to the large orbital moment and its $p$-type symmetry[31].

In the WHH theory[32], the SOC enters in two ways. On the one hand, as the SOC constant $\lambda_{SO} = \hbar / 3\pi\tau_{SO}k_B T_c$, where $\tau_{SO}$ is the characteristic scattering time due to the SOC. An increase in this constant for compounds with a strong SOC increases $\mu_0 H_{c2}(0)$. But this effect may be overpowered by the Fermi velocity increase, which affects the magnitude of the critical field through the Maki parameter $\alpha$:

$$\alpha = \frac{3\hbar}{2m\tau V_F^2} = \frac{3\hbar}{2ml_eV_F} = -\,0.527\left(\frac{d\mu_0 H_{c2}}{dT}\right)\Big|_{T=T_c}, \qquad (1)$$

where $\tau$ is the characteristic time of electron scattering on defects and impurities, $m$ is the electron mass, $l_e$ is the electron mean free path, and $V_F$ is the Fermi velocity[32]. The key role here is played by the Fermi velocity, which for polyhydrides usually takes the value 2.5 - 3.8×$10^5$ m/s[33]. However, this is only true for those cases where the contribution of hydrogen dominates at the Fermi level. In the case of lower hydrides such as $P2_1/m$-$BiH_2$, the main contribution comes from bismuth atoms. Calculation with accounting the SOC based on the band structure of $BiH_2$ at 155 GPa[34] gives the Fermi velocity of about 11×$10^5$ m/s, 3-5 times higher than in most other hydride superconductors. A similar situation is observed for another superhydride, cubic $SnH_4$: the average Fermi velocity reaches 10.4×$10^5$ m/s at 200 GPa (Supporting Table S5). Thus, with the same quality of sample microcrystals (i.e., value of $l_e$), $P2_1/m$-$BiH_2$ will have a 3-5 times smaller derivative $\mu_0 H_{c2}/T_c$ and, accordingly, a smaller $H_{c2}(0)$ within the WHH model, which is in good agreement with our experiment.

**The critical current density of $BiH_2$**

We investigated the critical current density in the DAC's B2 sample after its degradation with possible decomposition of $BiH_4$. Evidence for this process includes the appearance of a strong Raman signal of hydrogen (Supporting Figure S7), as well as the disappearance of a pronounced transition with an onset of about 80-90 K in the temperature dependence of electrical resistance (Figure 2c). An important property of superconductivity in $BiH_2$, as in $SnH_4$[10], is the possibility of almost complete suppression of the superconductivity in this compound using the magnetic field of the current flowing through the sample. To do this, it is necessary to use a high current of up to 0.14-0.16 A (Figure 4b, Supporting Figure S12a). In the constant current mode, such measurements cannot be performed due to the burnout of the connecting wires in the high-pressure diamond cell, so we used a rectangular current pulse amplifier and a Keithley 6221 current source (Figure 4a). As a result, $V$-$I$ characteristics were obtained and measured in the range from 45 to 10 K (1st experiment, $J_c(0) = 1\times10^{10}$ A/m$^2$) and down to 2 K (2nd experiment, $J_c(0) = 4.44\times10^9$ A/m$^2$) with a $T$-step of 1 K (Figure 4b, Supporting Figure S12). There is a small leakage current in the system, which is expressed in the negative initial slope of the superconducting part of the $V$-$I$ characteristics. Using the criterion $V(I_c) > 2$ μV[35, 36] (as well as 1, 3 μV, see Supporting Figure S11) allows us to calculate the critical current, which reaches 60 mA at 2 K, and the lower bound of critical current density $J_c(0) = 1\times10$ A/m$^2$ for the maximum cross-section of the near-electrode space of 2×7 μm$^2$.

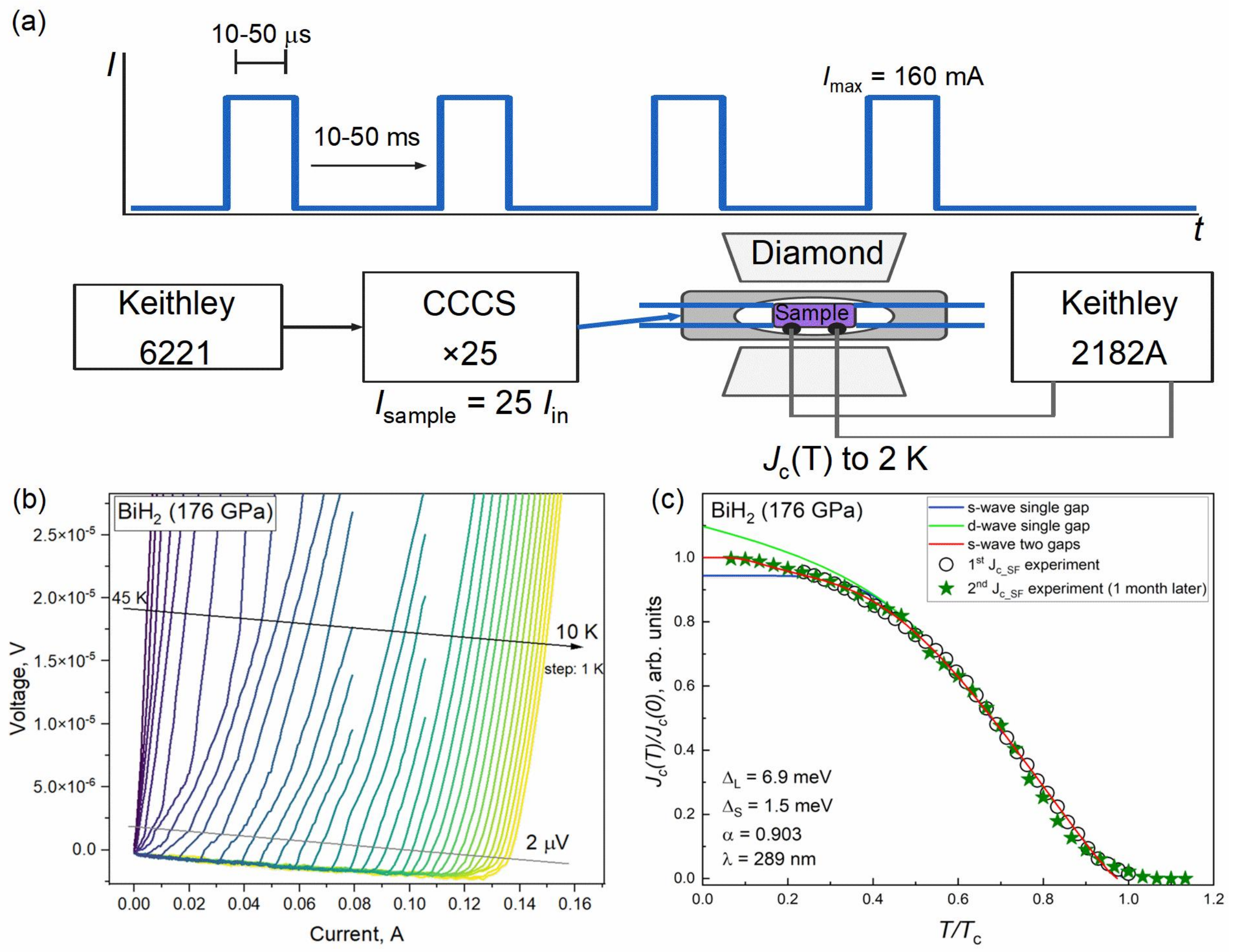

FIG. 4. The critical current density of $BiH_2$. (a) Schematic of the pulsed-current measurement setup. The programmed current from a Keithley 6221 is amplified by a custom current-controlled current source with a gain of 25 and applied to the $BiH_2$ sample in a four-probe DAC configuration. Rectangular pulses with durations of 10-50 μs and intervals of 10-50 ms enable peak currents up to 160 mA while reducing time-averaged Joule heating. The voltage is recorded using a Keithley 2182A, allowing ($J_c$(T)) measurements down to 2 K. (b) Voltage-current characteristics of DAC B2 sample at 176 GPa before degradation (the 1st experiment) in the temperature range from 45 K to 10 K with a step of 1 K. The critical current for each temperature was determined when the voltage exceeded 2 μV-line above the background. (c) Temperature dependence of the critical current density for two experiments performed with a one-month interval (black open circles - before degradation; green stars - after degradation of $T_c$(offset)). The data were interpolated using three models: two-gap *s*-wave (red line), single gap *s*-wave (blue line), and single gap *d*-wave (green line) superconductivity. The two-gap *s*-wave model yielded the best results.

The shape of the $J_c(T)$ curve with a large number of experimental points also allows us to establish the model (for instance, *s*-wave or *d*-wave) of superconductivity realized in the compound. The lowest temperature region is especially important for this purpose. For most superhydrides, this region is practically impossible to study directly due to the excessively high current (1-10 A) that must be applied to suppress superconductivity near 0 K. It is only possible to indirectly determine the critical current near 0 K from the magnetic moment of the sample, detected using a SQUID magnetometer[37]. Applying the Talantsev-Tallon model[14, 15], which is a development of the Carrington model[38], and which we tested earlier on the example of $SnH_4$[10], we found that $BiH_2$ exhibits *s*-wave superconductivity, which can be best described only by including a small additional gap $\Delta_s(0)$ = 1.5 meV in addition to the main gap $\Delta_L(0)$ = 6.9 meV. This explains the absence of pronounced saturation of $J_c(T)$ at low temperature. The limitation of this approach is in the heterogeneity of the $BiH_2$ sample and the unknown exact geometric parameters of the "weak link" through which the critical current density flows.

The possible presence of multigap superconductivity in hydrides has been previously discussed in a number of papers[39-41]. In X-H systems with a heavy atom X (for instance, m(Bi)/m(H) ≈ 200 for the Bi-H system), there are two independent groups of phonons of different characteristic energies that practically do not interact directly. Accordingly, the Eliashberg function, the ratio of which is approximately equal to the ratio of the characteristic phonon frequencies: $\omega_{Bi}/\omega_H \approx$ 5-6 in the case of $BiH_2$. Detailed calculations based on the fully anisotropic Migdal-Eliashberg equations, performed using the EPW package[42-44] for $P2_1/m$-$BiH_2$ (the most likely phase responsible for the observed superconductivity), suggest, however, that the superconducting gap is not of two-gap character, but rather highly anisotropic (see later discussion and Figure 5).

## Discussions

To elucidate the gap structure of the $P2_1/m$-$BiH_2$ phase, we solved the fully anisotropic Migdal-Eliashberg equations using maximally localized Wannier functions, as implemented in the EPW code[42-44]. Since the electronic density of states is nearly constant in the vicinity of the Fermi level, we employed the Fermi-surface-restricted approximation[45]. The main results are summarized in Figure 5, where we show calculations performed at a pressure of 176 GPa.

As expected, the Bi phonons are confined to a narrow low-frequency region, approximately between 0 and 30 meV, whereas the H phonon modes extend over a much broader energy range, reaching nearly 400 meV. This yields a total electron-phonon coupling parameter of $\lambda$=1.26 and a logarithmic phonon frequency $\omega_{log}$ = 44 meV, as shown in Figure 5a.

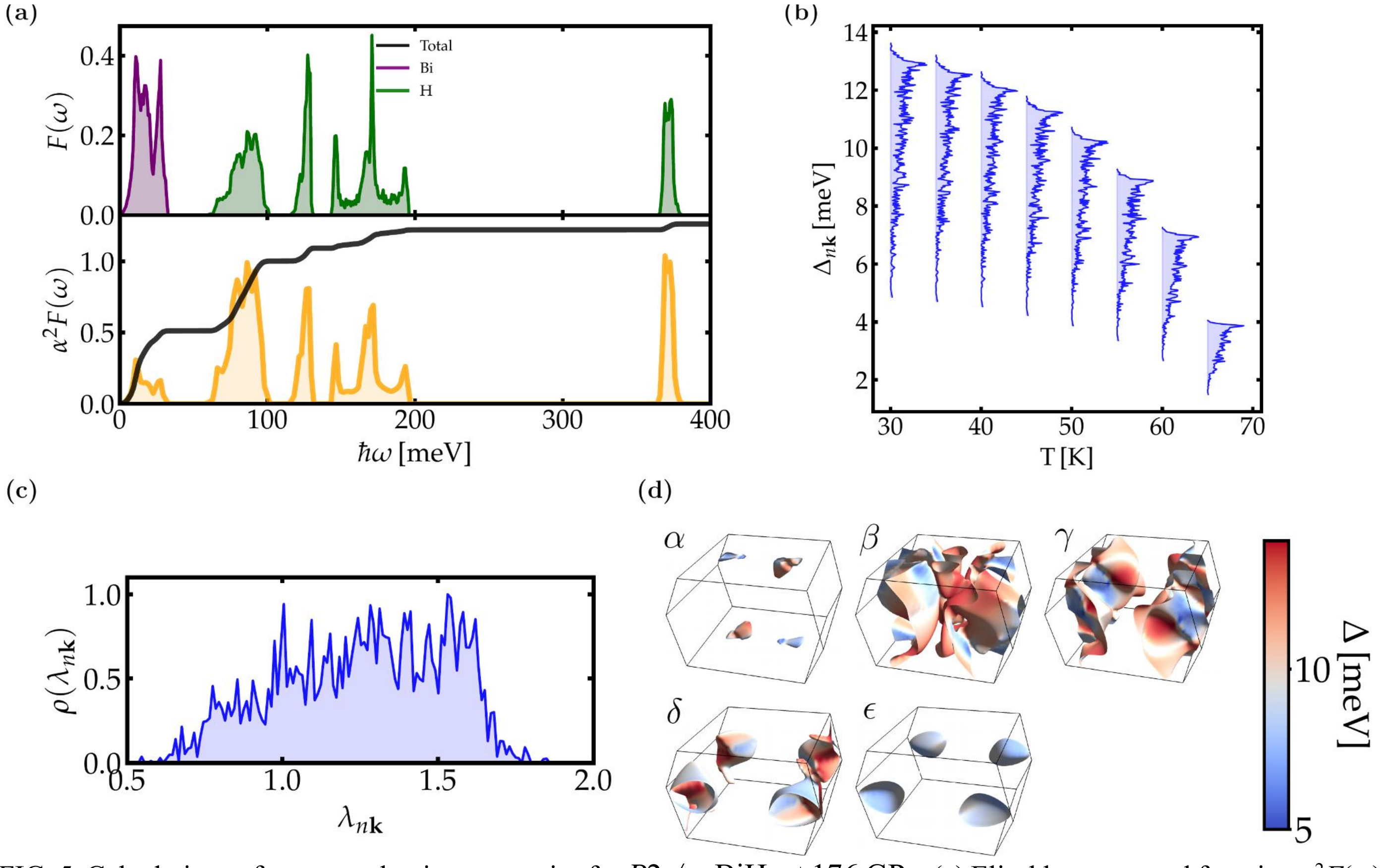

FIG. 5. Calculations of superconducting properties for $P2_1/m$-$BiH_2$ at 176 GPa. (a) Eliashberg spectral function $\alpha^2F(\omega)$ (orange solid lines), cumulative total el-ph coupling parameter $\lambda$ (black solid lines) and phonon density of states $F(\omega)$. (b) Anisotropic gap distribution on the Fermi surface as a function of temperature. (c) Distribution of the total el-ph coupling strength. (d) Momentum-resolved superconducting gap on the Fermi surface calculated within the ab initio Migdal-Eliashberg theory for $T$ = 30 K.

Figure 5b shows the distribution of the anisotropic superconducting gap values, $\Delta_{nk}$, on the Fermi surface (FS) as a function of temperature. Rather than forming well-separated gap distributions, as expected for genuinely multiband superconductors such as $MgB_2$[44], $\Delta_{nk}$ displays a broad and continuous distribution. At 30 K, $\Delta_{nk}$ ranges from approximately 5 to 14 meV. This behavior is consistent with the distribution of the electron-phonon coupling strength over the FS, shown in Figure. 5c, where $\lambda$ also exhibits a broad and continuous spread, ranging from about 0.5 to nearly 1.9. The gap closes at approximately 67 K, in good agreement with our experimental results.

To further examine the gap structure, we analyzed the momentum resolved superconducting gap on the FS, as shown in Fig. 5d. Five distinct bands cross the Fermi level ($\alpha$, $\beta$, $\gamma$, $\delta$ and $\varepsilon$), each displaying a broad and continuous distribution of $\Delta_{nk}$. Notably, the $\varepsilon$ sheet shows comparatively lower gap values ranging from 5 to approximately 10 meV, while the remaining sheets exhibit significantly larger gaps. However, since $\varepsilon$ constitutes only a small fraction of the total FS weight, $P2_1$/m-$BiH_2$ is best described as a single but strongly anisotropic superconductor rather than a genuine two gap system. This situation is reminiscent of $NbSe_2$[46], where a smaller gap resides on the FS sheets around Γ and a larger, highly anisotropic gap dominates the sheets centered at K. There too, the gap distributions overlap within the same energy window at any given temperature, producing experimental signatures, including two shoulders in STM spectra, that closely resemble a pristine two gap response, even though the underlying physics is that of a strongly anisotropic gap.

In this light, the two-gap fit of the $J_c(T)$ data presented above, while yielding a good description of the data, should not be regarded as definitive evidence for two distinct order parameters. The critical current density is an indirect probe of the superconducting state, strongly affected by vortex pinning, weak links, disorder, sample geometry, and flux creep, and does not directly measure the momentum dependence of $\Delta_{nk}$. A single but strongly anisotropic gap, with minima and maxima on the FS, can produce an average temperature dependence that mimics the response expected from two separate gaps. We therefore conclude that the most consistent interpretation of both our experimental and theoretical results is that $P2_1/m$-$BiH_2$ a strongly anisotropic single gap superconductor, and that the apparent two gap signatures arise from this anisotropy rather than from the presence of two distinct order parameters.

In summary, we synthesized superconducting $BiH_2$ at 157-176 GPa and observed transition temperatures of 58-70 K. Its comparatively low upper critical field of 11-17 T corresponds to a relatively long coherence length and, together with an experimentally accessible critical current, makes $BiH_2$ a suitable platform for current-dependent studies under megabar pressure. By using short rectangular current pulses, we reduced sustained Joule heating and applied currents up to 160 mA, enabling measurements of the critical current down to 2 K. The resulting $J_c$(T) exhibits two characteristic energy scales, described phenomenologically by effective gap values of approximately 6.9 and 1.5 meV. Fully anisotropic Migdal-Eliashberg calculations instead reveal a single, nodeless gap with pronounced momentum dependence, suggesting that the apparent two-gap response originates from strong gap anisotropy rather than two independent superconducting gaps. Unlike tunnelling spectroscopy, this approach does not require a dedicated tunnel junction and can be implemented using conventional DAC transport configurations. These results demonstrate that pulsed critical-current measurements provide a practical gap-sensitive transport approach for investigating hydrides and other superconductors under extreme pressure.

## Acknowledgements

This work was supported by the National Key R&D Program of China (Grant No. 2022YFA1405500), and National Natural Science Foundation of China Youth Basic Research Program for Doctoral Students (No. 125B2071). V.V.S. acknowledges the financial support from Shanghai Science and Technology Committee, China (No. 22JC1410300) and Shanghai Key Laboratory of Materials Frontier Research in Extreme Environments, China (No. 22dz2260800). D. S. and D. Z. thank EMFL-ISABEL project for financial support of this research. P.N.F. acknowledges support from the Austrian Science Fund (FWF) under project DOI 10.55776/ESP8588124. We would like to thank the staff of the BL10XU (Spring-8, Japan) beamline and, especially, Dr. Hirokazu Kadobayashi, for the assistance in carrying out this study as part of project 2025A1058. This work was also supported by HLD-HZDR, member of the European Magnetic Field Laboratory (EMFL).

X.H., D.S., and V.S. led the project. J.G., D.S., P.N.F, S.L., T.H., C.H. and X.H. wrote the manuscript. J.G., D.S., and I.T. prepared the high-pressure experiments. J.G., D.S., D.Z., S.L., T.H., and X.H. performed the electrical transport measurements. Y.W., Y.C., S.C., K.Z., and X.W. processed the XRD experiments. A.S.U., L.A.M., and V.M.P. performed some of the theoretical calculations. P.N.F performed the EPW calculations for $BiH_2$. J.G., D.S., and I.T. contributed equally to this work.

# SUPPORTING INFORMATION

# Superconducting gap in covalent bismuth dihydride under extreme conditions

Jianning Guo[1,†], Dmitrii V. Semenok[2,*,†], Ivan A. Troyan[3,†], Pedro N. Ferreira[4], Di Zhou[2], Yulong Wang[1], Yuzhi Chen[1], Su Chen[1], Kexin Zhang[1], Xinyue Wu[1], Sven Luther[5], Toni Helm[5], Andrey V. Sadakov[6], Alexander T. Daniyarhodzhaev[6], Alexey S. Usoltsev[6], Leonid A. Morgun[6], Vladimir M. Pudalov[6], Christoph Heil[4], Viktor V. Struzhkin[3,7,*], and Xiaoli Huang[1,*]

[1] *State Key Laboratory of Superhard Materials, College of Physics, Jilin University, Changchun 130012, China*

[2] *Center for High Pressure Science and Technology Advanced Research (HPSTAR), Beijing, China*

[3] *Shanghai Key Laboratory of Material Frontiers Research in Extreme Environments (MFree), Shanghai Advanced Research in Physical Sciences (SHARPS), Pudong, Shanghai 201203, China*

[4] *Institute of Theoretical and Computational Physics, Graz University of Technology, NAWI Graz, 8010, Graz, Austria*

[5] *Hochfeld-Magnetlabor Dresden (HLD-EMFL) and Würzburg-Dresden Cluster of Excellence, Helmholtz-Zentrum Dresden-Rossendorf (HZDR), Dresden 01328, Germany*

[6] *V. L. Ginzburg Center for High-Temperature Superconductivity and Quantum Materials, P. N. Lebedev Physical Institute, Russian Academy of Sciences, Moscow 119991, Russia*

[7] *Center for High Pressure Science and Technology Advanced Research, Shanghai 201203, China*

* Corresponding authors, emails: dmitrii.semenok@hpstar.ac.cn (D. V. Semenok), viktor.struzhkin@hpstar.ac.cn (V. V. Struzhkin), huangxiaoli@jlu.edu.cn (X. Huang)

# 1. Methods

### Sample synthesis and characterization

The synthesis of the $BiH_x$ samples was accomplished by multiple Nd:YAG fiber IR laser heating pulses (1064 nm, pulse length: 200-300 ms, spot size: 3-5 μm) of the Bi sample in a sublimated ammonia borane ($NH_3BH_3$) environment at the target pressure. The Raman laser (532 nm) spot size was about 2-3 μm.

Crystal structure of Bi polyhydrides synthesized in DAC B1 was studied using the synchrotron X-ray diffraction (XRD) on the BL15U1 beamline with a beam of wavelength of 0.6199 Å at the Shanghai Synchrotron Research Facility (SSRF). The synchrotron X-ray diffraction spot was larger: for SSRF it was 15×8 $\mu m^2$, and for SPring-8 it was 3×4 $\mu m^2$. X-ray diffraction patterns from the DAC B2 sample were collected on the BL10XU beamline with a beam of wavelength of 0.4143 Å at the SPring-8 (Japan). The sequence of experiments was for DAC B1: transport studies → XRD → pulsed field measurements; while for DAC B2: transport studies → steady magnetic field measurements → V-I ($1^{st}$ exp.) → V-I ($2^{nd}$ exp) → XRD

Resistance measurements in steady magnetic fields were performed using delta mode (DAC B1, Keithley 6221 current source) or/and low-frequency AC mode ($f$ = 313.3 Hz, SR830 lock-in) with the RMS current of 0.1 mA (DAC B2). In all cases, a four-pin van der Pauw scheme was used.

Resistance measurements in a pulsed magnetic field were performed in a four-contact van der Pauw scheme using an alternating current of 2 mA with a frequency of 7.77 kHz and 3.33 kHz. To analyze the magnetic phase diagram, we used the real part of the impedance ($Z = U/I$). The duration of the magnetic field pulse was 150 ms, and the maximum field amplitude was 50 Tesla. At the typical magnetic pulse field amplitudes up to 30 T, we did not observe any significant heating by eddy currents. The magnetic induction decay mode (downsweep) was used for the analysis. To control the temperature, we used a Cernox thermometer, glued to the DAC's B1 gasket with thermoconductive glue. To stabilize the sample temperature, an external heater, an insulated Ni-Cr wire of 1 m long and 50 microns in diameter with a resistance of about 100 Ohms, was used.

### Theoretical calculations

The phonon band structure and density of phonon states were computed using the Phonopy[1] package implementing the finite displacement method. The same package was used to visualize the phonon density of states and the band structure. K-paths were generated with the help of the VASPKIT. Supercells of $Bi_4H_8$ were generated for *Cmcm* and $P1$-$BiH_2$ and used for calculations of electron and phonon density of states. The energy cutoff and k-spacing parameters for the VASP structural optimization were set at 600 eV and $2\pi \times 0.1$ Å$^{-1}$, respectively. To calculate phonon frequencies and electron–phonon coupling (EPC) coefficients, we used the Quantum Espresso (QE) package[2] utilizing density functional perturbation theory (DFPT)[3], plane-wave pseudopotential method, and the PBE KJPAW pseudopotentials[4, 5]. The q-meshes for each structure were 4 × 2 × 4, k-mesh was 16 × 12 × 16. We used the optimized tetrahedron method[6] to calculate the Eliashberg function, and the Allen–Dynes formula[7] with the Coulomb pseudopotential $\mu^* = 0.1$ to estimate $T_c$ of bismuth hydrides.

## Theoretical calculations for $P2_1/m$-$BiH_2$

The electronic-structure calculations for $BiH_2$ were performed within density functional theory (DFT)[8, 9], using the Quantum ESPRESSO (QE) suite[2, 10]. The electron–ion interaction was represented by scalar-relativistic optimized norm-conserving Vanderbilt (ONCV) pseudopotentials[11, 12] taken from the PseudoDojo library[13]. Exchange–correlation effects were treated within the generalized-gradient approximation (GGA) using the Perdew–Burke–Ernzerhof (PBE) functional[14]. Plane-wave kinetic-energy cutoffs of 80 Ry were used. The convergence threshold for the self-consistent-field (SCF) cycle was set to $10^{-10}$ Ry.

Brillouin-zone integrations were carried out using Γ-centered Monkhorst–Pack k-point meshes[15]. In particular, we used a 12 × 12 × 12 mesh for $BiH_2$. Metallic states were described with Methfessel–Paxton smearing[16] with widths of 0.02 Ry. These settings ensured convergence of the total energies within 1 meV/atom. Both lattice parameters and internal atomic positions were relaxed until the variation in total energy was smaller than $10^{-7}$ Ry and the residual forces were below $10^{-6}$ Ry/$a_0$.

Phonon spectra and electron–phonon matrix elements were obtained within density-functional perturbation theory (DFPT)[3]. For the phonon calculations, the self-consistency threshold was tightened to $10^{-16}$ Ry. Dynamical matrices and electron–phonon matrix elements were computed on uniform q-point meshes of 6 × 6 × 6.

Maximally localized Wannier functions were generated using the Wannier90[17] code through the automated projection scheme in combination with the selected columns of the density matrix (SCDM) method[18, 19]. We retained 15 Wannier functions in the disentangled subspace, excluding the lowest 10 bands from the Wannierization procedure. The SCDM disentanglement was performed using an erfc-type window, with $\mu$ = 24.5 eV and $\sigma$ = 3.0 eV, while the upper bound of the frozen disentanglement window was set to 31 eV. The Wannierization was allowed to run for up to 3000 iterations. The interpolation was based on a 12 × 12 × 12 coarse electronic k-point mesh and a 6 × 6 × 6 coarse phonon q-point mesh. The Fermi energy was fixed to 24.5368 eV.

The fully anisotropic Migdal–Eliashberg equations were solved with the EPW code. Electron energies, phonon frequencies, and electron–phonon matrix elements were interpolated on fine k- and q-point grids containing $36^3$ points. The minimum phonon frequency included in the calculations was set to 10 $cm^{-1}$. A Fermi window width of 1 eV was used. The Matsubara-frequency cutoff was set to $\omega_{max}$ = 1 eV. The broadenings used for the energy-conserving delta function and for the q-space summation in the electron–phonon coupling were, respectively, 0.1 eV and 0.1 meV. The Migdal–Eliashberg equations were solved using a Coulomb pseudopotential of $\mu^*$ = 0.111, which, for $BiH_2$, corresponds to $\mu^*$ = 0.1 within the Allen-Dynes formalism. The superconducting gap obtained on the imaginary axis was analytically continued to the real axis using the Nevanlinna scheme[20].

**Table S1.** Parameters of samples, high-pressure DACs, and diamond anvils used in this study. All DACs used in the experiment are asymmetrical custom-designed DACs. We used HPHT type IIa diamond anvils. The electrodes were created by magnetron sputtering of Mo (DAC B1), or tantalum and gold (Ta/Au) through a mask in the case of DAC B2. The thickness of electrodes doesn't exceed 0.3-0.5 μm.

| | **DAC B1** (DAC for pulsed fields, 12 mm in diameter) | **DAC B2** (Troyan's design DAC, 25 mm in diameter) |
|---|---|---|
| **Starting material** | Bi/AB | Bi/AB |
| **DAC's material** | NiCrAl | BeCu |
| **Insulating gasket** | W/BN/epoxy | nonmagnetic steel/$CaF_2$/epoxy |
| **Electrodes** | Mo | Ta/Au |
| **Culet diameter, μm** | 50 | 50 |
| **Pressure** | 157-163 GPa | 169-176 GPa |

| Composition | Bi + $BiH_{2+x}$ | Bi+$BiH_2$+$BiH_4$ |
|---|---|---|
| **Transition temperature (onset)** | 65-69 K | 90 K (before degradation)<br>57 K (after degradation) |
| **Performed experiments** | Transport measurements in pulsed and steady (0-8 T) magnetic fields, XRD | Transport measurements in steady fields (0-16 T), V-I characteristics, XRD |

Classifying hydrides according to the standard definition of type I and II superconductors is almost never possible. This requires calculating the Ginzburg-Landau parameter, $\kappa = \lambda_L/\xi$, which means knowing the London penetration depth and coherence length for the superconductor. While $\xi$ can be obtained from the experimental upper critical field at moderate values ($B_{c2}(0)$ < 60 T), measuring the penetration depth for microsamples squeezed between diamond anvils is currently impossible. $\lambda_L$ can be indirectly estimated from the lower critical field ($B_{c1}$) or the heat capacity jump, but measuring these parameters also faces significant experimental difficulties, and methods for their measurement have not yet been developed. Another approach, using various models (e.g., Talantsev's model) for the critical current density $J_c(T)$, runs into the problem of measuring the effective thickness of the superconducting phase in synthesized hydride samples. Thus, in this work we prefer to use the terms "soft" and "hard" superconductivity due to the lack of the possibility of calculating the Ginzburg-Landau parameter.

**Table S2.** Properties of two classes of superconducting hydrides. For both types $T_c$ may reach 200 K.

| **"Soft" superhydrides** | **"Hard" superhydrides** |
|---|---|
| Molecular hydrogen sublattice | Atomic hydrogen sublattice |
| Low concentration of defects and pinning centers | High concentration of defects and pinning centers |
| Relatively large microcrystals (1-10 μm), "spotty" point-like XRD patterns | Fine powder X-ray diffraction patterns (0.1-1 μm) |
| Low symmetry of the crystal structure<br>Partially disordered hydrogen sublattice | High symmetry of the crystal structure |
| Low thermodynamic stability | High thermodynamic stability<br>Effective synthesis via high-T laser heating |
| Low $H_{c2}(0)$, and $J_c(0)$ | High $H_{c2}(0)$, and $J_c(0)$ |
| $N_F$ is due to the contribution of electrons from a heavy atom, as a result, $V_F$ can be higher than in the case of clathrate atomic superhydrides | $N_F$ is mainly due to the contribution of *s*-electrons of hydrogen |

## 2. Transport measurements

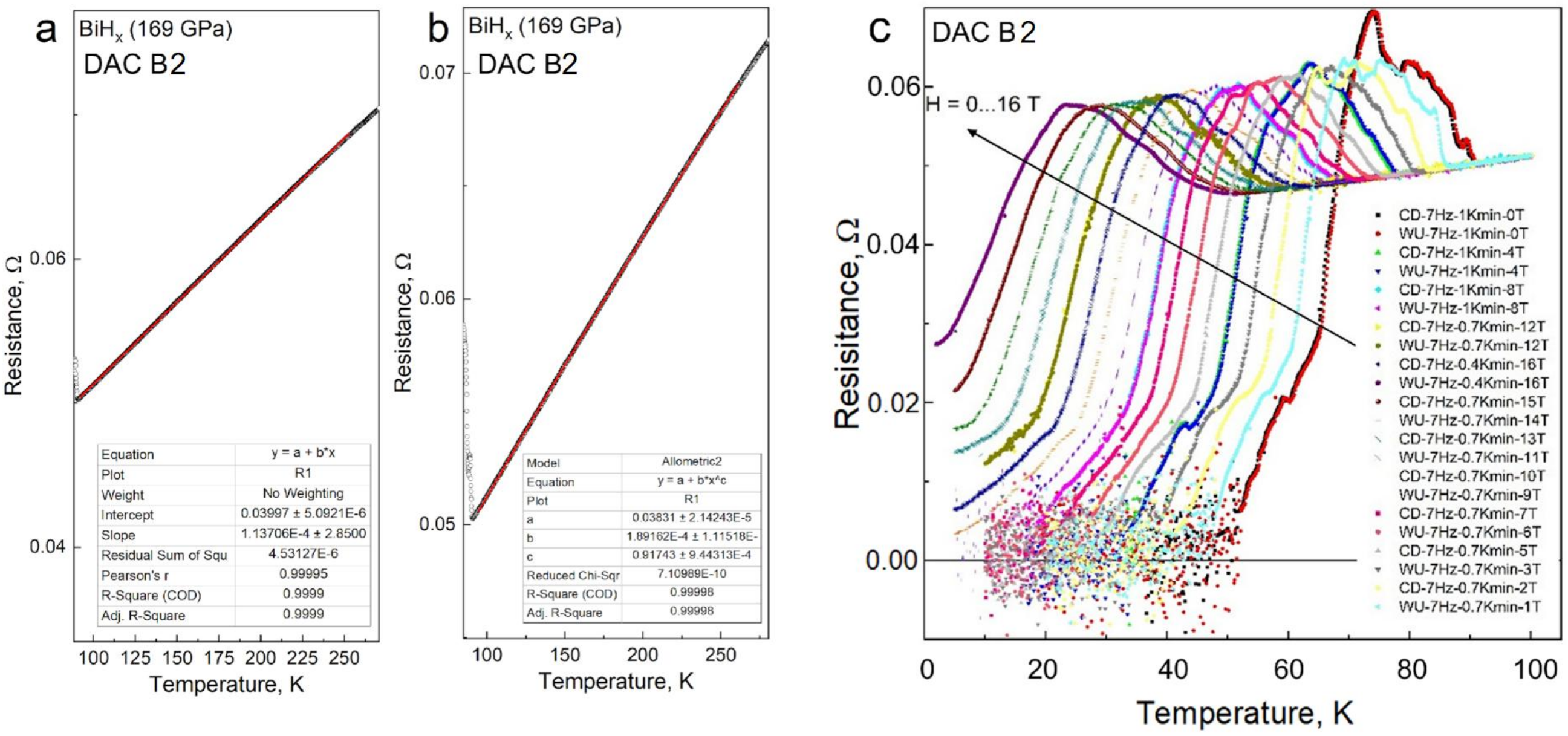

**Figure S1.** Temperature and magnetic field dependence of the electrical resistance of $BiH_4$ in the DAC B2 (before degradation). (a) Fit of the temperature dependence of the electrical resistance of DAC B2 in the non-superconducting state, performed using a linear function. (b) Fit of the temperature dependence of the electrical resistance of DAC B2 performed using a power-law function. (c) Electrical transport study of $BiH_x$ sample synthesized in DAC B2 in steady magnetic fields. The onset $T_c$ shifts to lower temperature in applied fields of 0–16 T. Measurements were done in the AC mode, the current frequency was 7 Hz, cooling/warming rates are 0.4, 0.7, and 1 K/min.

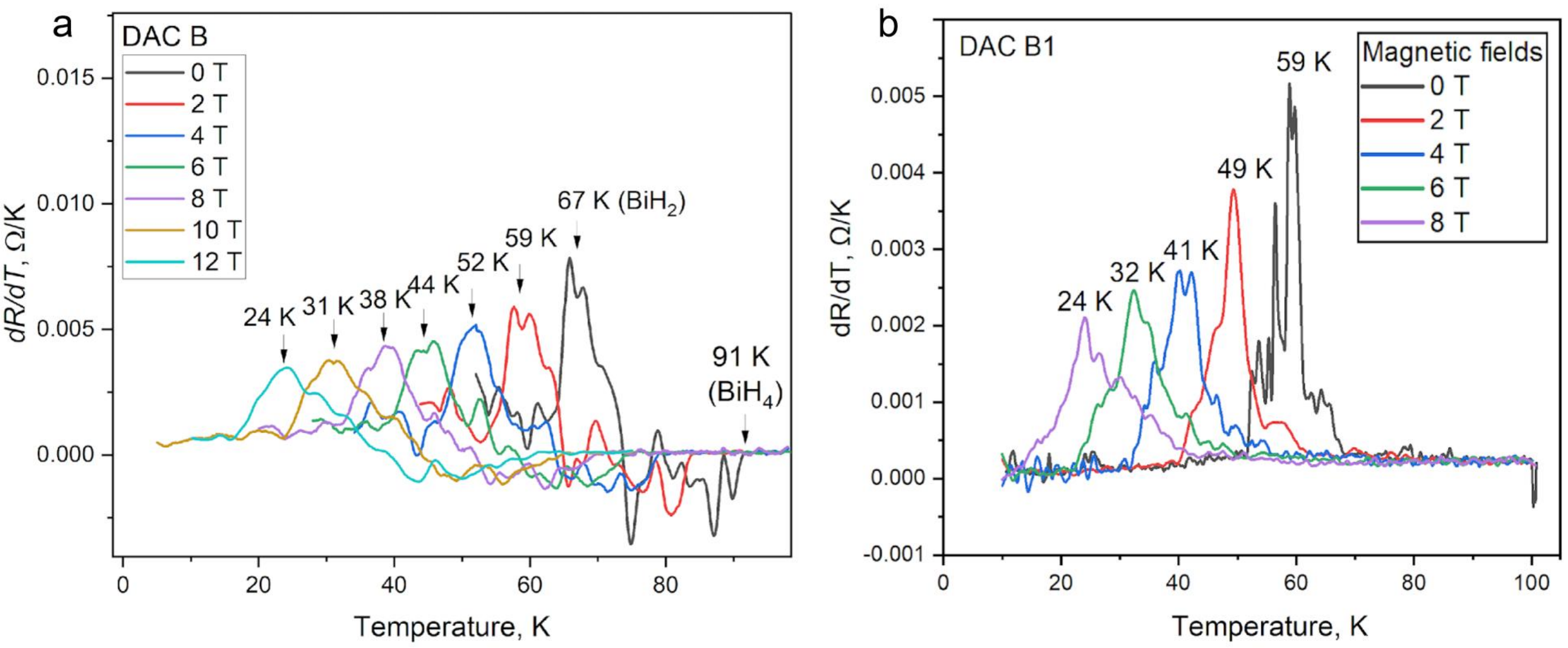

**Figure S2.** (a) Scheme for determining $T_c(BiH_2)$ in DAC B2 sample by the maximum of the *dR/dT* in different magnetic fields. (b) The same for DAC B1.

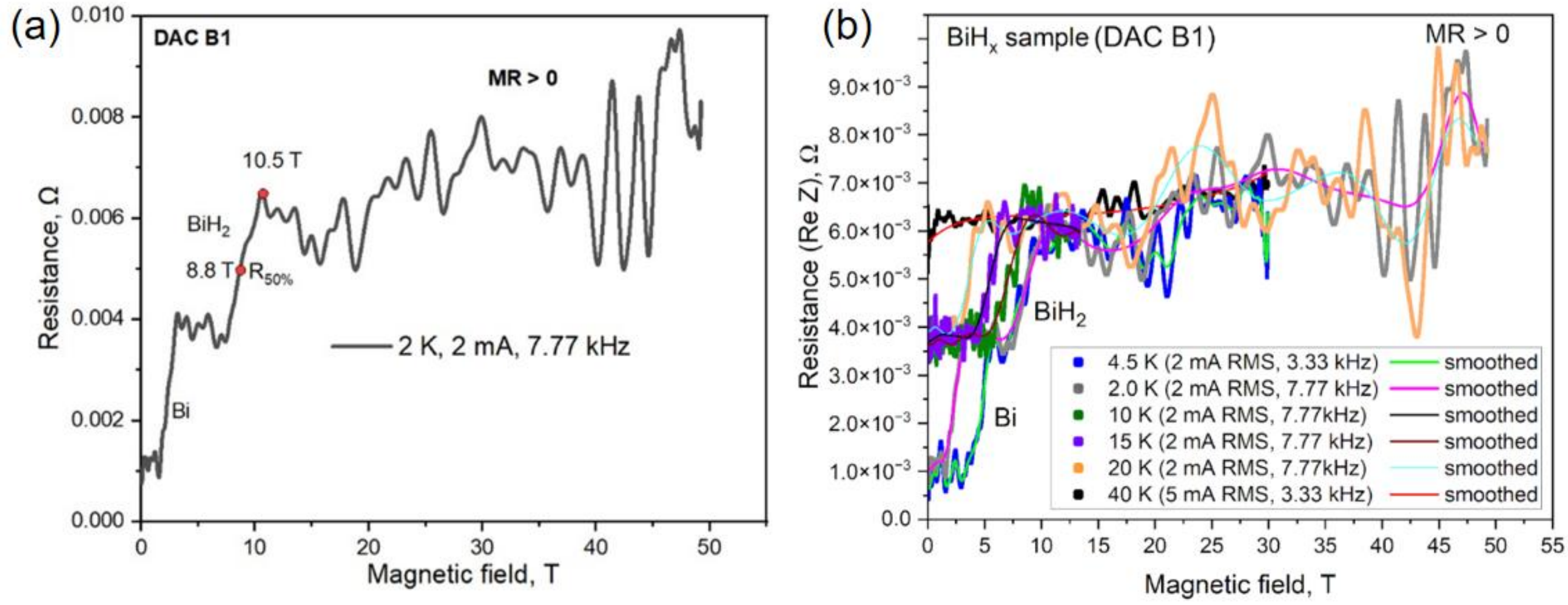

**Figure S3.** Extended data of the pulsed field measurements. (a) The raw R-H data at 2 K. Excitation current was 2 mA (RMS) to increase signal-to-noise ratio, but does not suppress superconductivity much. There are two steps on the R-H plot at 2 K, which probably correspond to the presence of metallic Bi (0-3 T) and $BiH_2$ (7 – 10.5 T) in the sample. For Bi we found $\mu_0 H_{c2}$(2 K) = 3 – 3.5 T, and for $BiH_2$ it is 10.5 T. (b) Raw and smoothed data for the dependence of electrical resistance (real part, Re) of the DAC B1 on magnetic field within pulse field experiments at different frequencies: 3.33 kHz and 7.77 kHz.

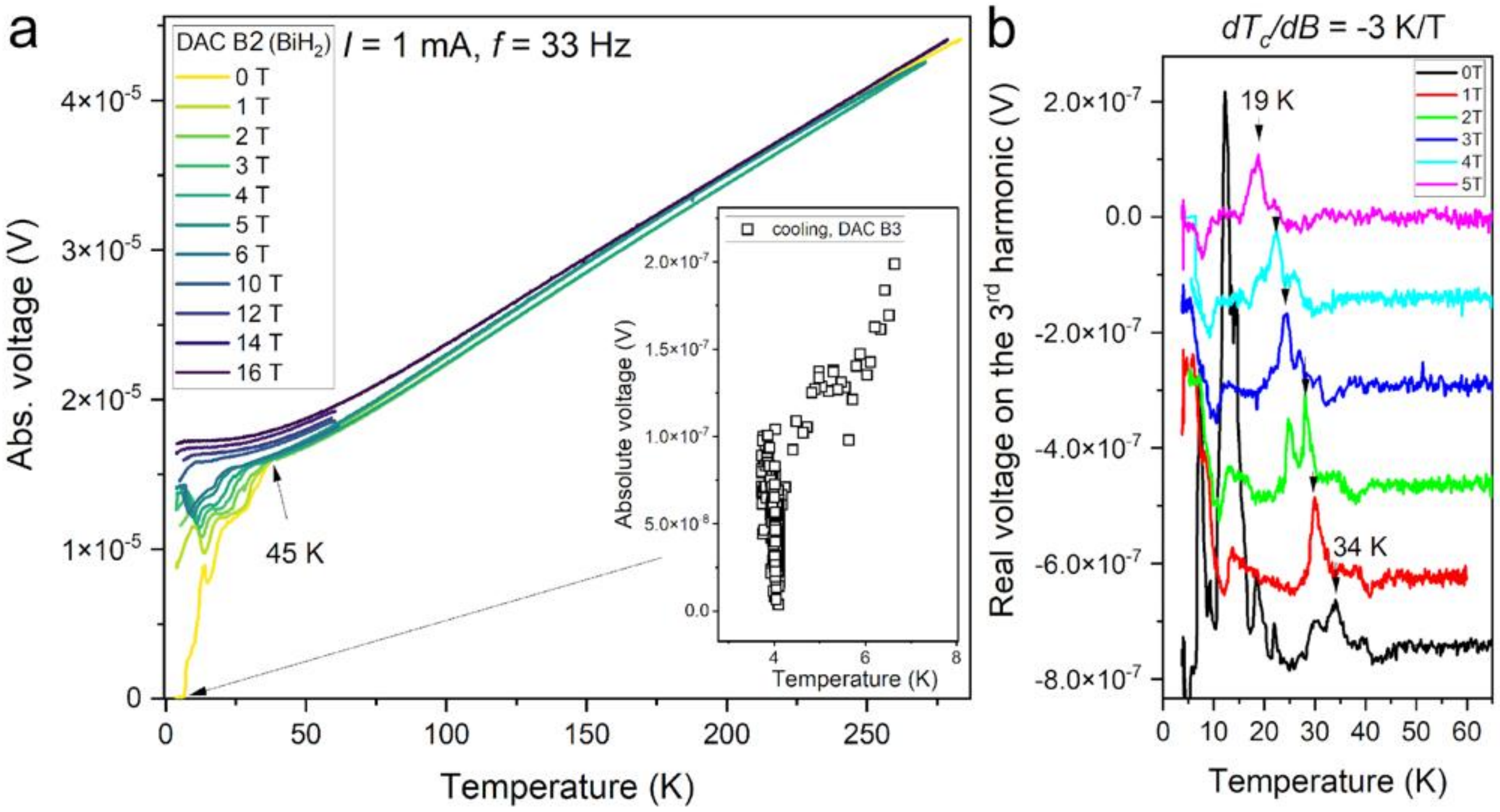

**Figure S4.** Behavior of bismuth hydride $BiH_2$ in a magnetic field after degradation of the DAC B2 sample (September 2025) and comparison of different hydrides by values of $T_c$ and $H_{c2}(0)$. (a) Temperature dependence of the voltage drop across the $BiH_2$ DAC B2 sample in magnetic fields from 0 to 16 T, measured in September 2025. It can be seen that traces of superconductivity practically disappear at 14-16 T. Inset: achievement of near-zero resistance in this sample at 4 K in the absence of a magnetic field. (b) Simultaneously detected dependence of the signal at the third harmonic (33×3= 99 Hz), which allows us to estimate $dT_c/dH \approx -3$ K/T from the position of the first peak of the 3rd harmonic.

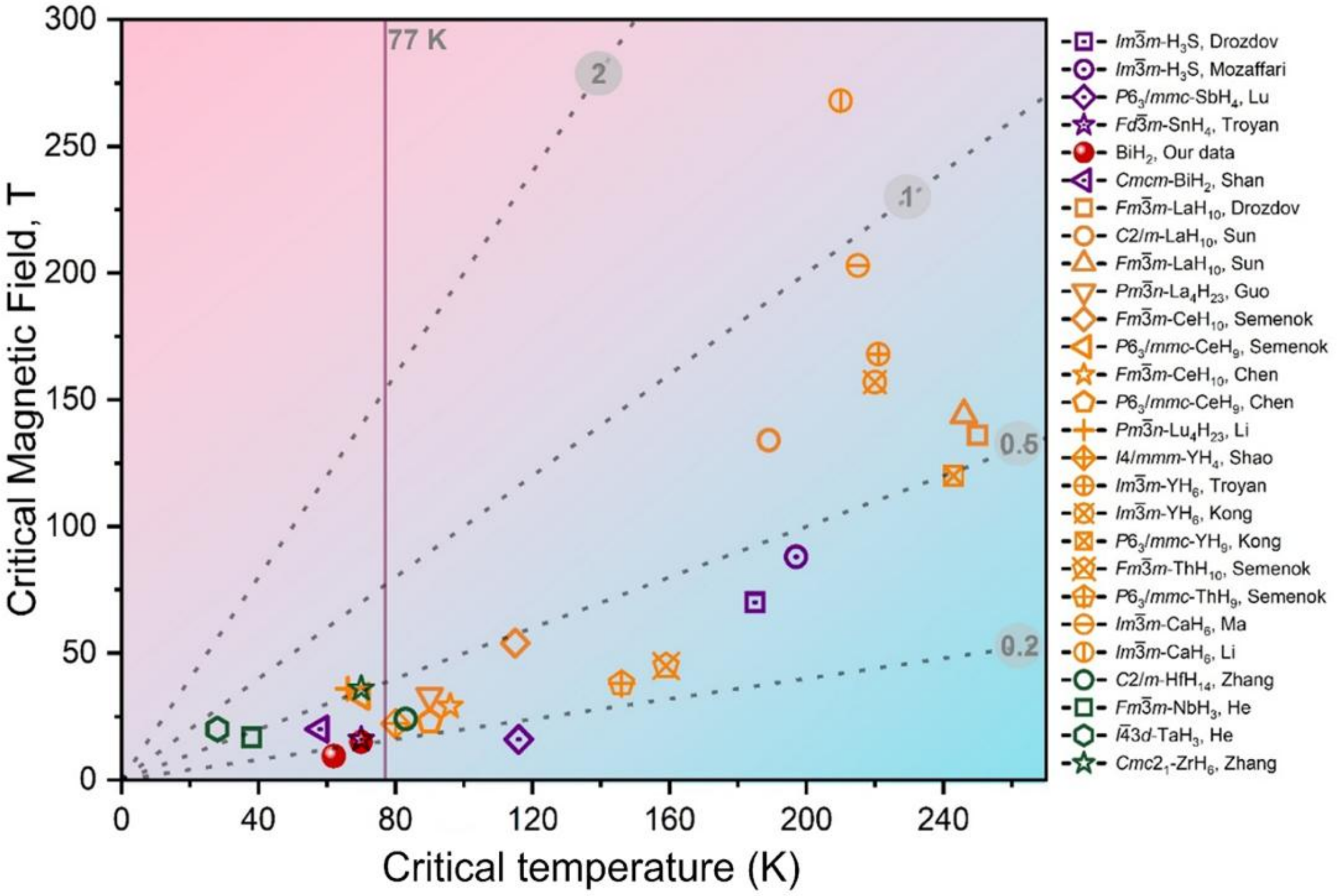

**Figure S5.** The relationship between the extrapolated upper critical magnetic field $\mu_0 H_{c2}(0)$ and the superconducting transition temperature ($T_c$) of various superconducting hydrides. The red solid circle represents this work. The orange, purple, and green symbols mark the cage-like hydrides, covalent hydrides, and transition metal hydrides, respectively.

## 3. Raman and XRD measurements

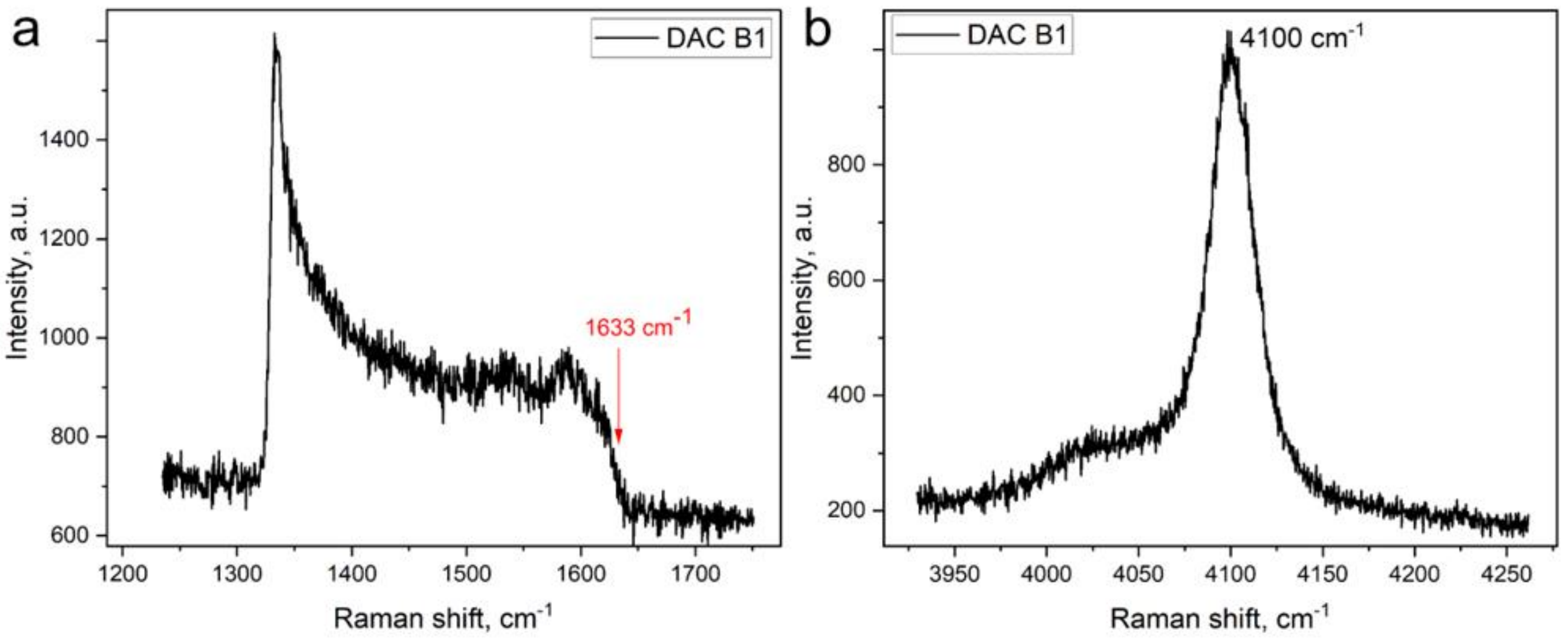

**Figure S6.** Raman studies of $BiH_2$ sample in DACs B1. (a,) Diamond edge region. (b) Hydrogen vibron detected in DAC B1. According to the frequency of $H_2$ oscillations, pressure in the DAC B1 is 140 GPa.

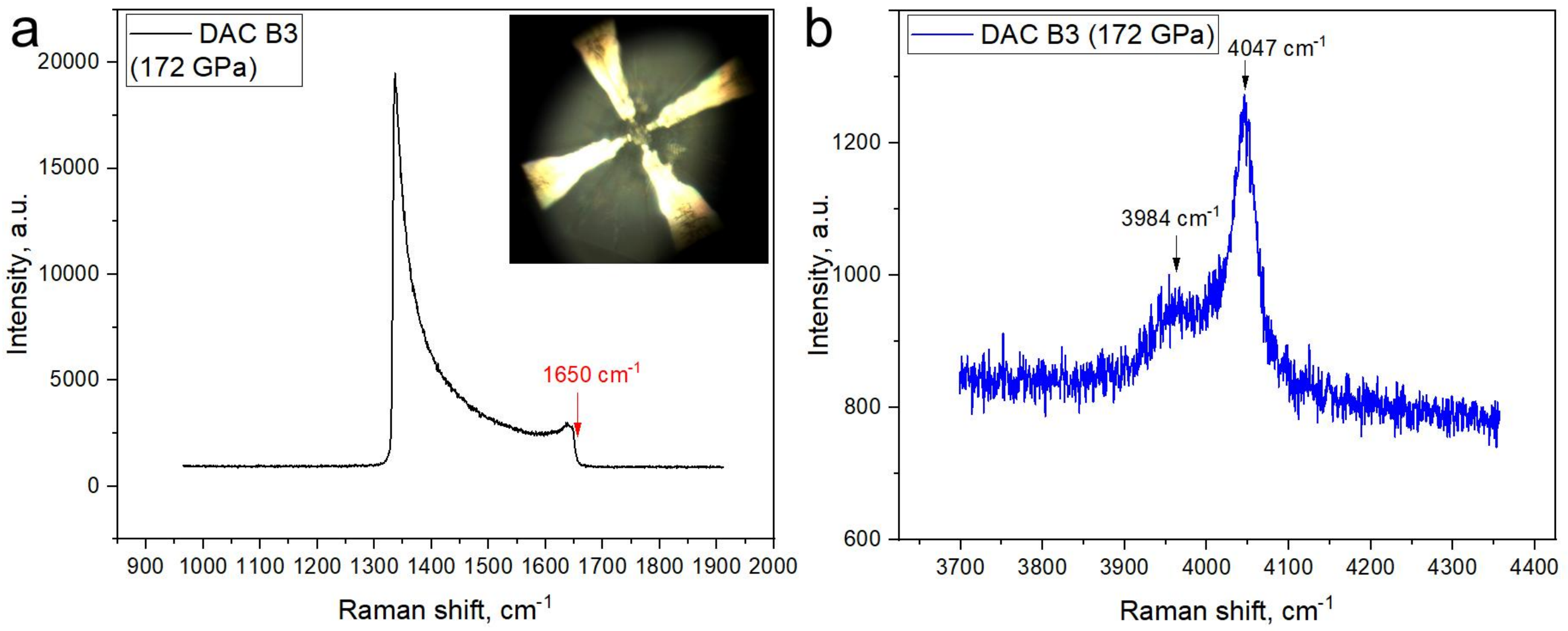

**Figure S7.** Raman spectra of the sample in DAC B2 at 172 GPa. (a) Raman spectrum showing the diamond Raman edge around 1650 cm⁻¹. The inset shows an optical image of the sample loaded in the diamond anvil cell. (b) Raman spectrum in the hydrogen region, showing peaks attributed to hydrogen vibrations (around 3984 cm⁻¹ and 4047 cm⁻¹).

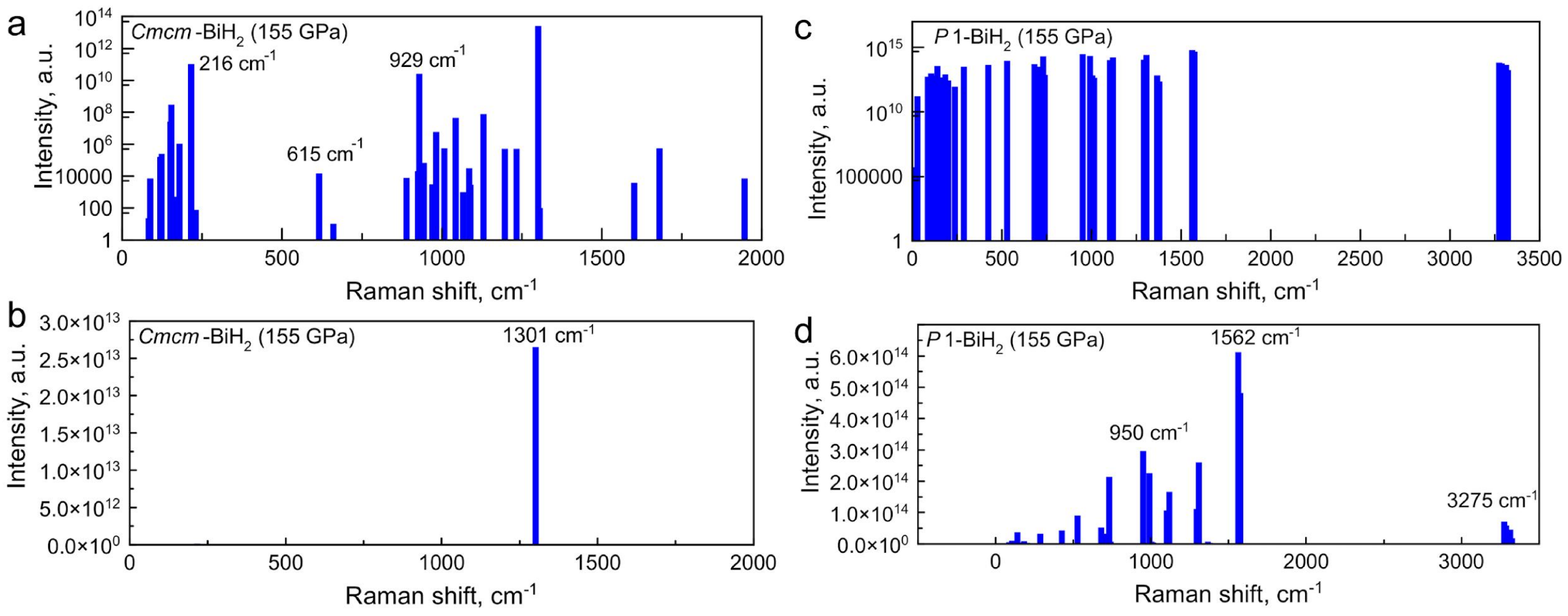

**Figure S8.** Calculated Raman spectrum of (a,b) *Cmcm*-$BiH_2$ and (c,d) *P*1 (pseudo $P2_12_12_1$)-$BiH_2$ at 155 GPa in (a, c) logarithmic, and (b, d) linear scales. We used a code (vasp_raman.py) designed for calculating off-resonance Raman spectra of dielectrics. Considering that Bi hydrides are metals, the intensities of individual spectral lines may not correspond to the calculated ones.

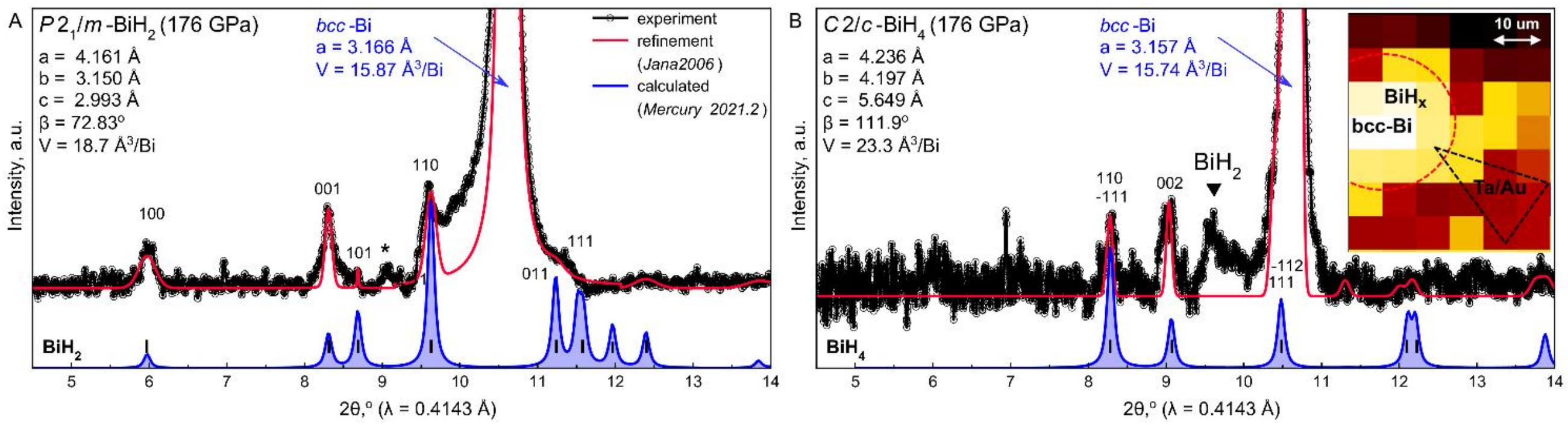

**Figure S9.** X-ray diffraction patterns of the DAC B2 samples at 176 GPa. The main reflections from the sample correspond to the presence of the $P2_1/m$-$BiH_2$, $C2/m$-$BiH_2$, and $C2/c$-$BiH_4$ phases. The $C2/m$-$BiH_2$ has a crystal

structure similar to the *Cmcm*. (a) Experimental pattern (black circles), Le Bail refinement of the unit cell parameters (red line), and the predicted XRD pattern (blue line). The diffraction peak 002 from $BiH_4$ is indicated by an asterisk. (b) The same for the $BiH_4$-enriched region of the sample. Inset: distribution map of bismuth and bismuth hydrides in a sample based on diffraction data. The map is a processing of 6×7 = 42 XRD patterns, a fragment of a larger 11×11-step XRD map (60×60 $\mu m^2$, see screenshot below). The light color indicates a higher intensity of the bismuth diffraction peak, which is also noted by the caption "*bcc*-Bi." Diffraction peaks at 2theta < 10 deg. were also detected in the vicinity of the bismuth particle, corresponding to bismuth hydrides $BiH_2$ and $BiH_4$. On the map, this zone is indicated by a red dashed circle and the caption "$BiH_x$." Finally, the black triangle indicates the region containing the characteristic signal from *bcc*-Ta, the main component of Ta/Au electrodes. Of course, the signal is weak and limited by the diffraction angle.

# 4. Critical current measurements

To enhance our ability to measure critical current beyond the capability of the Keithley 6221, we engineered our custom current booster, or, in other words, a current-controlled current source (CCCS). This CCCS allowed us to source current into the sample up to 0.6 A and perform voltage sweeps up to 60 V, while preserving the Keithley 6221's setup speed and resolution. The design was implemented with precautions that ensure the Keithley 2182a can operate within its high-resolution range (Figure S10). The current pulse duration was typically 10-50 microseconds, and the interval between pulses was 10-50 milliseconds. Since bismuth hydride does not fill the entire volume of the sample B2 uniformly, the actual cross-sectional area of the current bridge is less than 2×7 $\mu m^2$. Thus, in our V-I measurements, we are coming to a lower bound for the critical current density.

The current source operates in such a way that during idle time, the sample is held at the supply potential. When a measurement pulse is applied, the sample's potential shifts abruptly, which introduces a small artifact into the measurement circuit. Furthermore, the current-carrying wires and sample contacts collectively possess a resistance on the order of 100 Ohms. At currents of approximately 100 mA, this results in a common-mode voltage ($V_{CM}$) of up to 10 V. Typically, modern integrated amplifiers have a common-mode rejection ratio (CMRR) of around $10^6$ (120 dB). Thus, $10V/10^6$ results in an offset of ~10 μV. The monotonic offset on the order of several μV observed with increasing current is attributed precisely to this effect and does not affect the accuracy of the critical current determination.

There are several techniques to probe the order parameter in superconductors under pressure. Some of them are based on tunnel effects[21], while others are based on the investigation of the temperature dependence of the superfluid density. The latter proved to be a reliable tool to determine not only the quantity and values of energy gaps, but also the symmetry of the order parameter – whether it has s-wave, d-wave, or even more complex structure[22]. This usually requires either measuring directly the London penetration depth $\lambda(T)$, or the lower critical field, or measuring the self-field critical current density $J_c(T)$. Given all the limits and difficulties of measurements in DAC under high pressure, we were only able to measure the self-field critical current.

Talantsev *et al* suggested that several superconducting parameters such as the energy gap and the London penetration depth can be derived from the temperature dependence of the self-field critical current density. The parameters obtain by this method for conventional superconductors[23], for copper oxide high temperature superconductors[22] and for iron based high temperature superconductors[24] are in a good agreement with available data.

For a type II superconductor with the rectangular cross section critical current density is given

by[25]

$$J_c(T) = \frac{\Phi_0(ln(\kappa) + 0.5)}{4\pi\mu_0\lambda^3(T)}\left(\frac{\lambda(T)}{a}tanh\left(\frac{a}{\lambda(T)}\right) + \frac{\lambda(T)}{b}tanh\left(\frac{b}{\lambda(T)}\right)\right), \quad (S1)$$

where $\mu_0$ is vacuum permeability, $\Phi_0$ is the flux quantum, $2a$ and $2b$ are the width and the height of the sample, respectively, $\lambda(T)$ is London penetration depth and $\kappa$ is Ginzburg-Landau parameter. The latter has a weak temperature dependence and thus $ln(\kappa)$ can be thought as being effectively constant. In the case of a cylindrical Fermi surface the temperature dependence of the normalized superfluid density is given by[25]

$$\rho_S(T) = 1 - \frac{1}{2\pi k_B T}\int_0^{2\pi} d\theta\, cos^2(\theta) \times \int_0^{\infty} cosh^{-2}\left(\frac{\sqrt{\varepsilon^2+\Delta^2(T,\theta)}}{2k_B T}\right) d\varepsilon, \quad (S2)$$

where $k_B$ is the Boltzmann constant and $\Delta(T,\theta)$ is an angle dependent gap function. For an isotropic Fermi surface the expression becomes

$$\rho_S(T) = \frac{\lambda^2(0)}{\lambda^2(T)} = 1 - \frac{1}{2k_B T}\int_0^{\infty} cosh^{-2}\left(\frac{\sqrt{\varepsilon^2 + \Delta^2(T)}}{2k_B T}\right) d\varepsilon. \quad (S3)$$

The temperature dependence of the superconducting gap can be approximated by the expression given by Gross *et al*[26]

$$\Delta(T) = \Delta(0) tanh\left(\frac{\pi k_B T_c}{\Delta_0}\sqrt{a(T_c/T - 1)}\right)$$

where $T_c$, $2\Delta(0)/k_B T_c$ and $a$ are fitting parameters. An angle dependent gap function $\Delta(T,\theta)$ can be represented $\Delta(T,\theta) = \Delta(T) \cdot g(\theta)$ with $\Delta(T)$ being the temperature dependence of the gap function and $g(\theta)$ the unity magnitude angular dependence. For an anisotropic s-wave case $g(\theta) = (1{+}m cos(4\theta))/(1{+}m)$ with $m$ being anisotropy parameter and for the d-wave pairing $g(\theta) = cos(2\theta)$.

The temperature dependence of the critical current density was obtained from current-voltage characteristics measured from 2 K up to $T_c$. Critical current was determined from the current voltage characteristic by the current when voltage on the sample reaches 2 μV which is a widely used criterion in the literature[27]. The sample cross section for the calculation of the critical current density was determined from the geometrical parameters.

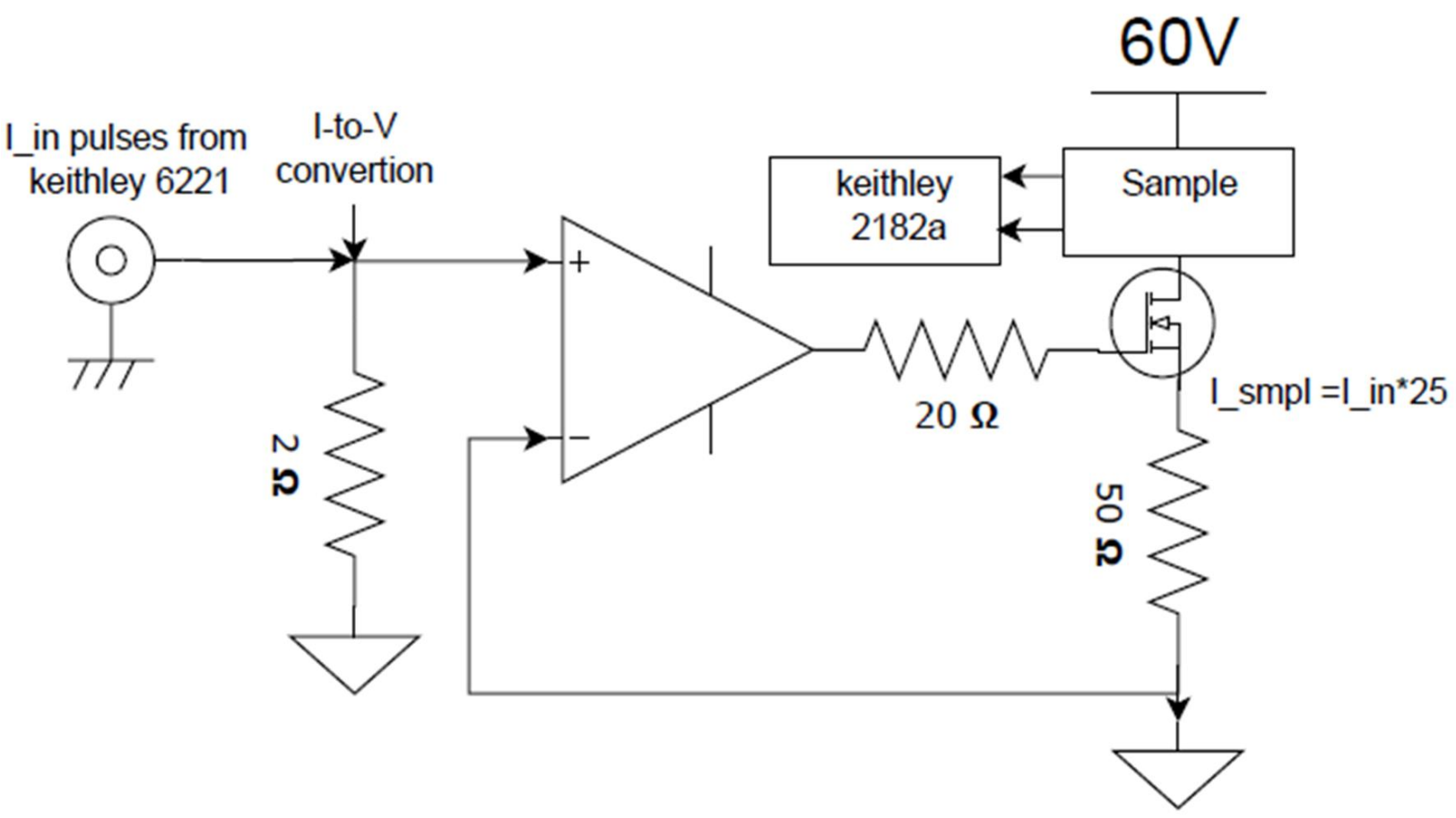

**Figure S10.** Scheme of measurements of voltage-current characteristics of the DAC B2 sample in the pulse mode using an amplifier of rectangular current pulses received from Keithley 6221.

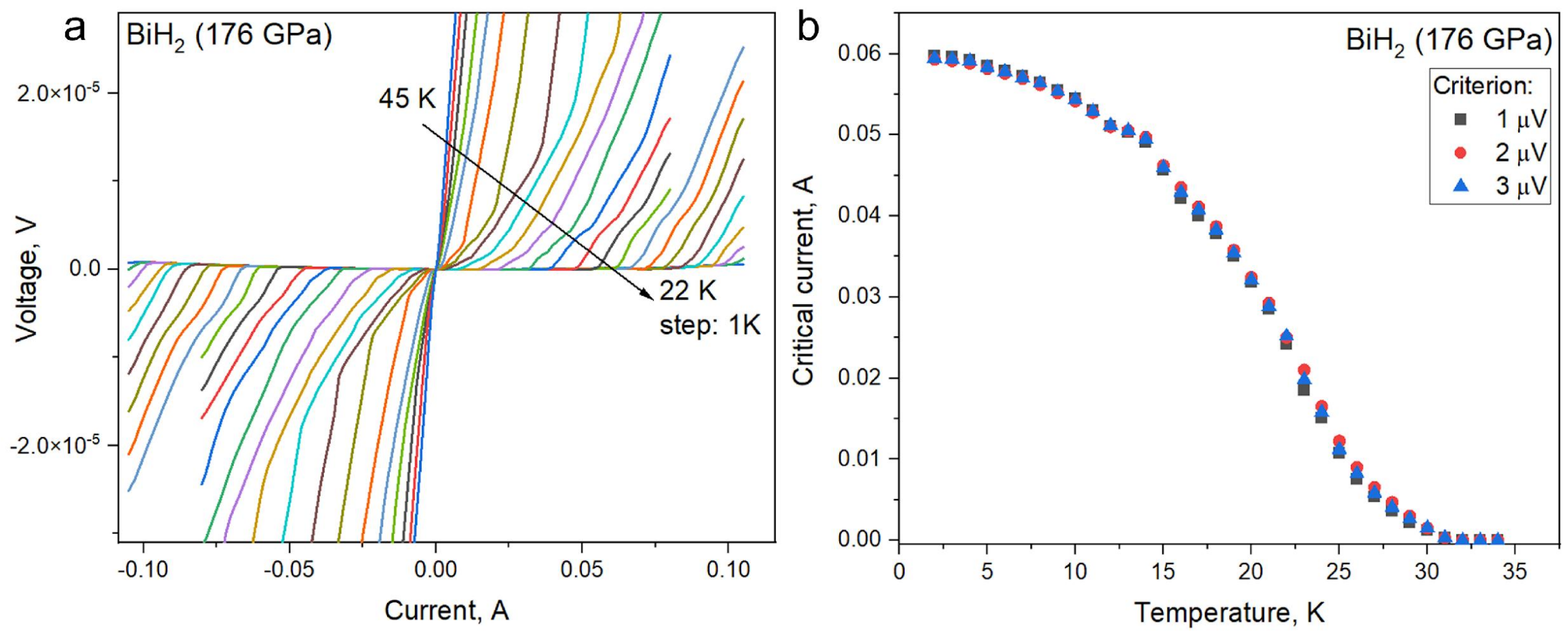

**Figure S11.** Voltage-current characteristics and temperature dependence of the critical current of $BiH_x$ sample in the DAC B2. (a) Voltage-current characteristics before degradation of the sample's $T_c$(offset). Measurements were performed in the temperature range from 45 K to 22 K with a 1 K step. (b) Temperature dependence of the critical current after degradation of the sample's $T_c$(offset). The critical current is determined using different critical voltage criteria: 1 μV, 2 μV, and 3 μV above the background.

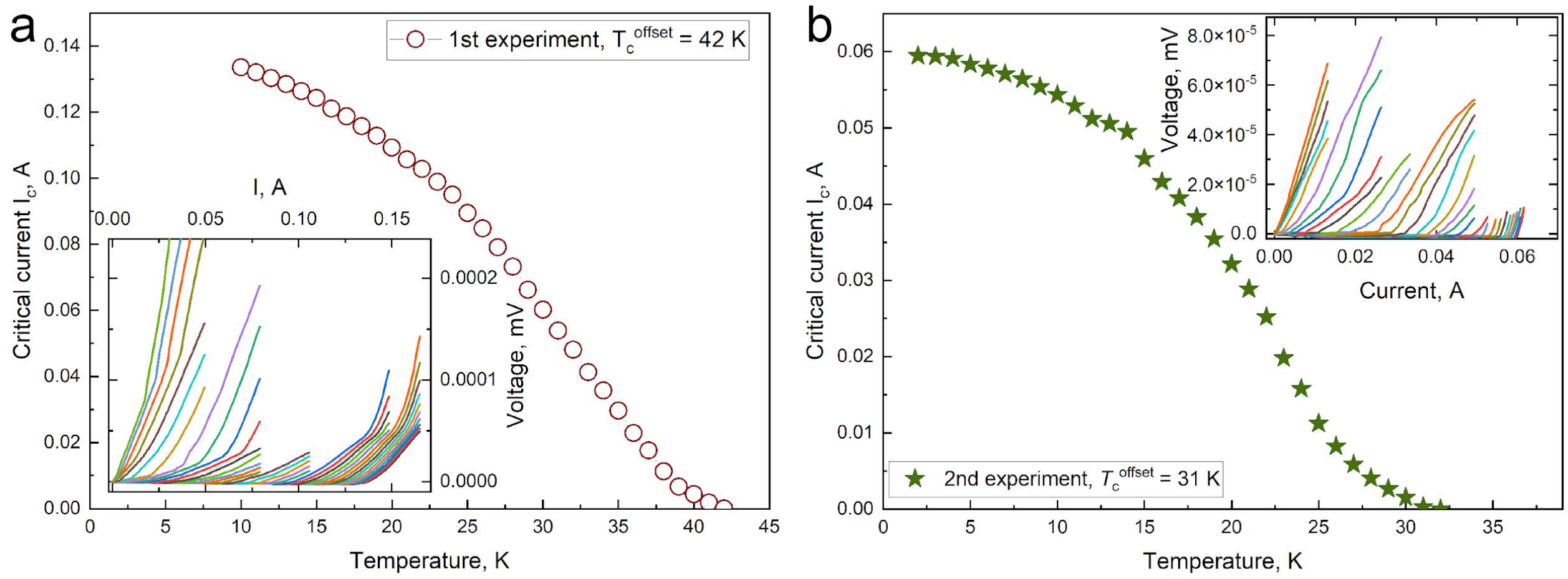

**Figure S12.** Temperature dependencies of the critical current in DAC B2. Two series of experiments: (a) before degradation of $T_c$(offset) from 42 K to 31 K, and (b) after this degradation down to 2 K.

## 5. Theoretical calculations

A structural search for stable bismuth polyhydrides is beyond the scope of the present study and has been addressed in previous theoretical works[28, 29]; here, we focus on discussing the stability and superconductivity of these already known phases. When we move from metal hydrides to non-metal ones (e.g. Bi, Sn, Se, P, As), we encounter their relatively low thermodynamic stability. In the reaction $Bi + nH_2 \rightarrow BiH_{2n}$, the equilibrium favors the formation of bismuth hydrides only at very high pressures of about 150 GPa, and the depth of the convex hull for the Bi-H system does not exceed 100 meV/atom at 200 GPa[28]. For comparison, in the Y-H system at the same pressure, the convex hull depth is about 1 eV/atom, 10 times greater. The key challenge for our work is to achieve a synthesis of pure phases. Our two experimental attempts (DACs B1-B2) to obtain superconducting bismuth hydrides demonstrate the difficulties discussed above. Only in the third attempt (DAC B2) were we able to achieve the fully superconducting state. The XRD study shows the formation of several bismuth hydrides and the presence of a significant amount of unreacted bismuth despite the presence of excess hydrogen. It is likely that Bi hydrides are thermally unstable and can dissociate into their constituent elements - bismuth and hydrogen - during laser heating.

There are several variants of the hydrogen sublattice in $BiH_2$: (1) a higher-symmetry *Cmcm* with hydrogen chains (Figure S21a, inset), and (2) lower-symmetry $P2_1/m$, $P2_12_12_1$, and $P1$ structures differing in the degree of orientation of the array of hydrogen molecules. In all cases, the bismuth sublattices are almost the same (*Cmcm*) and give rise to a similar XRD pattern. Thus, they cannot be well distinguished via XRD experiments.

Calculations show that *Cmcm*-$BiH_2$ should have the highest $T_c$ of about 85 K at 155 GPa (Figure S21a), but intrinsic infinite H-chains of this structure are dynamically unstable and decompose with the formation of $H_2$ molecules (Supporting Figure S13). This is particularly evident from molecular dynamics simulations at 300 K: unlike polymeric nitrogen[30], where infinite chains and all sorts of cycles are a widespread phenomenon, hydrogen at a pressure of 1-2 Mbar prefers to form separate molecules. Dynamically and thermodynamically stable $P2_1/m$-$BiH_2$ was predicted in 2015[28] to be a superconductor with $T_c$(150 GPa) = 59 K, which is in good agreement with the experimental data and our own estimates from fully anisotropic Migdal-Eliashberg calculations (Figure 5). In this crystal structure, all hydrogen molecules are oriented in the same way. Further disordering of the H-sublattice would decrease the $T_c$ of $BiH_2$: for instance, for $P1$ (pseudo $P2_12_12_1$), we obtained $T_c$ (155 GPa) ≈ 34 K (Figure S21c). However, it differs significantly from the experimentally obtained value of 61-70 K. Moreover, all these low-symmetry crystalline modifications are dynamically unstable in the harmonic approximation (see Supporting Figures S10-S12). From the point of view of compliance with experimental data and the highest dynamical stability, the $P2_1/m$-$BiH_2$ phase appears to be the most plausible explanation, whereas the analysis of the behavior of low-symmetry $BiH_2$ phases requires the use of a more complex anharmonic approach[31], which is beyond the scope of this study. We see that the degree of ordering of $H_2$ molecules in the bismuth dihydride sublattice has a very large effect on the critical temperature of superconductivity. This factor may also help explain the strong deviation[32] between the theoretical $T_c$ for $BiH_4$ (125 K at $\mu$ = 0.1) and the experimental values (90-91 K). Furthermore, since several possible $BiH_2$ structures exhibit very similar XRD patterns, we calculated the Raman shifts for both Cmcm-$BiH_2$ and P1-$BiH_2$ to aid structural identification (Supporting Figure S8). However, no Raman peaks corresponding to these calculated wavenumbers were observed in our experiments.

All $BiH_2$ structures are metals with a dominant contribution from the *p*-orbitals of the Bi atoms and a minor contribution from the *s*-orbitals of hydrogen to the density of electron states at the Fermi level (Figure 4b, d). This explains the low $T_c$ of bismuth hydrides compared to, for example, sulfur trihydride $H_3S$[33]. It is important to note that the large contribution of Bi atoms to the density of states at the Fermi level causes a significant increase in the Fermi velocity up to $V_F = 1.1\times10^6$ m/s at 155 GPa compared to other polyhydrides (Supporting Table S4 and Figure S20). Raman spectra could help us identify the products of the reaction of bismuth with hydrogen. However, as calculations show (see Supporting Figure S8), the most significant Raman lines of metallic $BiH_2$ fall into the region of intense Raman signals from diamond, $NH_3BH_3$, and its thermal decomposition products. Therefore, a confident identification of weak Raman signals from metallic $BiH_x$ against a very strong background remains very challenging (Supporting Figures S5-S6).

**Table S3.** Structures of various bismuth hydrides used in this work.

| **Structure (pressure)** | **Atomic coordinates (CIF format)** |
|---|---|
| *Cmcm*-$BiH_2$ (155 GPa)<br>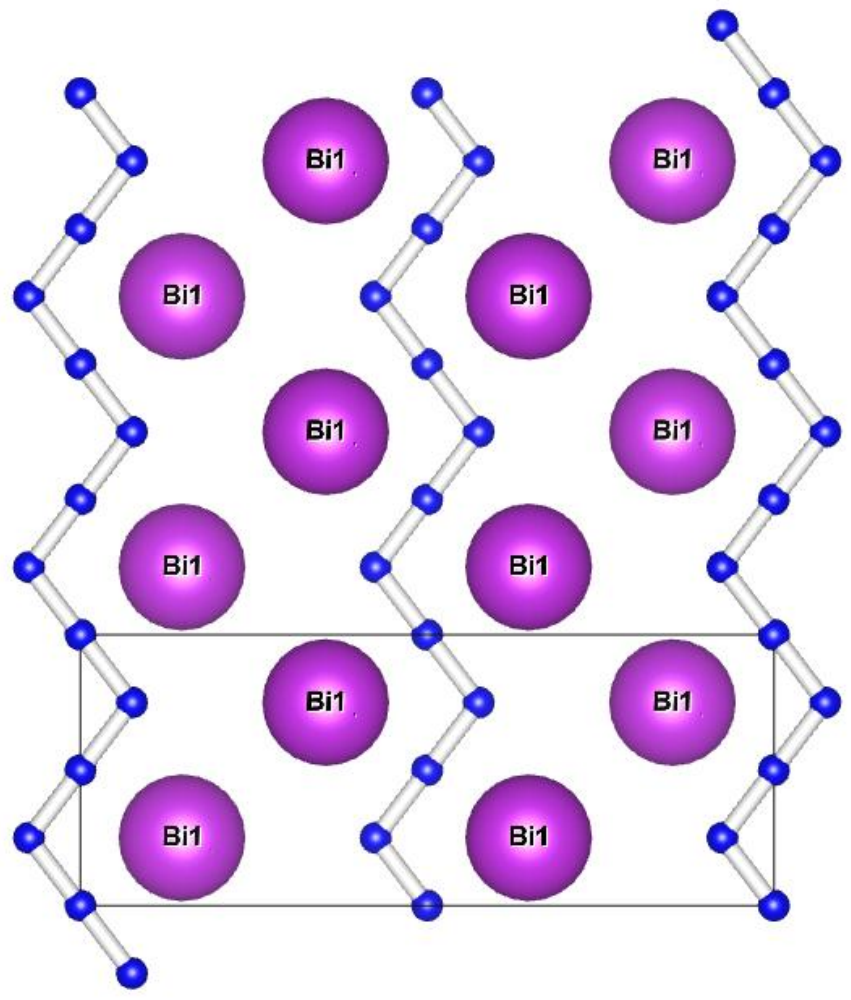<br><br>Space Group: *Cmcm* (#63-1)<br>a = 3.179 Å α = 90.0°<br>b = 7.961 Å β = 90.0°<br>c = 3.106 Å γ = 90.0°<br>V = 78.6067 Å$^3$ | _symmetry_space_group_name_H-M 'C 2/m 2/c 21/m'<br>_symmetry_Int_Tables_number 63<br>_cell_length_a 3.17900<br>_cell_length_b 7.96100<br>_cell_length_c 3.10600<br>_cell_angle_alpha 90.00000<br>_cell_angle_beta 90.00000<br>_cell_angle_gamma 90.00000<br><br>loop_<br>_space_group_symop_operation_xyz<br>x,y,z<br>x,-y,-z<br>-x,y,-z+1/2<br>-x,-y,z+1/2<br>-x,-y,-z<br>-x,y,z<br>x,-y,z+1/2<br>x,y,-z+1/2<br>x+1/2,y+1/2,z<br>x+1/2,-y+1/2,-z<br>-x+1/2,y+1/2,-z+1/2<br>-x+1/2,-y+1/2,z+1/2<br>-x+1/2,-y+1/2,-z<br>-x+1/2,y+1/2,z<br>x+1/2,-y+1/2,z+1/2<br>x+1/2,y+1/2,-z+1/2<br><br>loop_<br>_atom_site_label<br>_atom_site_type_symbol<br>_atom_site_fract_x<br>_atom_site_fract_y<br>_atom_site_fract_z<br>_atom_site_occupancy<br>Bi1 Bi 0.00000 -0.35367 0.25000 1.00000<br>H1 H 0.00000 -0.07420 0.25000 1.00000<br>H2 H 0.00000 0.00000 0.00000 1.00000 |
| *Pnma*-$BiH_2$ (125 GPa) | data_Pnma-BiH2 125 GPa<br>_audit_creation_date 2024-10-14<br>_audit_creation_method 'Materials Studio'<br>_symmetry_space_group_name_H-M 'PNMA'<br>_symmetry_Int_Tables_number 62 |

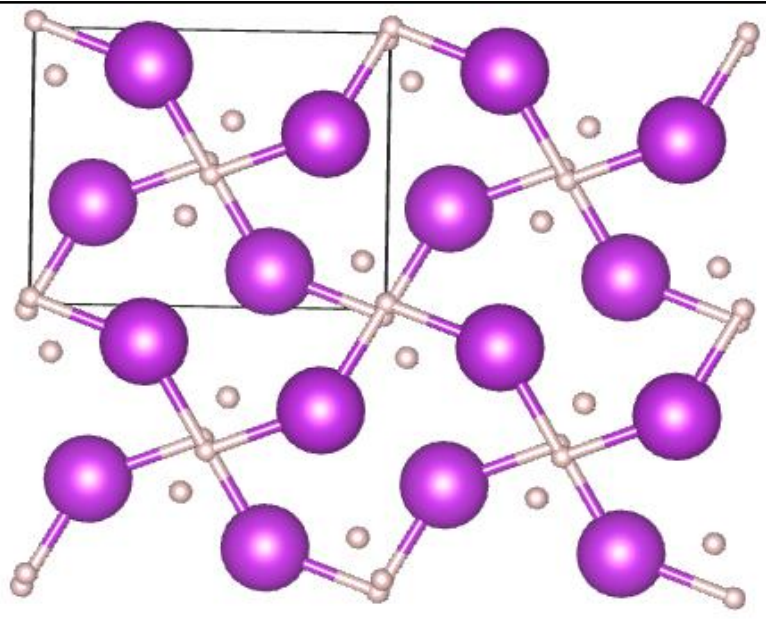

Space Group: *Pnma* (#62-1)
a = 5.905 Å  α = 90.0°
b = 3.159 Å  β = 90.0°
c = 4.577 Å  γ = 90.0°
V = 85.37 $Å^3$

```
_symmetry_cell_setting           orthorhombic
loop_
_symmetry_equiv_pos_as_xyz
  x,y,z
  -x+1/2,-y,z+1/2
  -x,y+1/2,-z
  x+1/2,-y+1/2,-z+1/2
  -x,-y,-z
  x+1/2,y,-z+1/2
  x,-y+1/2,z
  -x+1/2,y+1/2,z+1/2
_cell_length_a                   5.9050
_cell_length_b                   3.1590
_cell_length_c                   4.5770
_cell_angle_alpha                90.0000
_cell_angle_beta                 90.0000
_cell_angle_gamma                90.0000
_cell_volume                     85.3789
loop_
_atom_site_label
_atom_site_type_symbol
_atom_site_fract_x
_atom_site_fract_y
_atom_site_fract_z
_atom_site_U_iso_or_equiv
_atom_site_adp_type
_atom_site_occupancy
 H1   H    0.49300  0.25000  0.47800  0.00000  Uiso  1.00
 H2   H    0.06300  0.25000  0.17300  0.00000  Uiso  1.00
 Bi   Bi   0.17400  0.25000  0.63100  0.00000  Uiso  1.00
```

*P1 (pseudo Pnnm)-$BiH_2$* (155 GPa)

Symmetrization with tolerance 0.2 transforms this structure to *Pnnm*-$BiH_2$ (155 GPa)

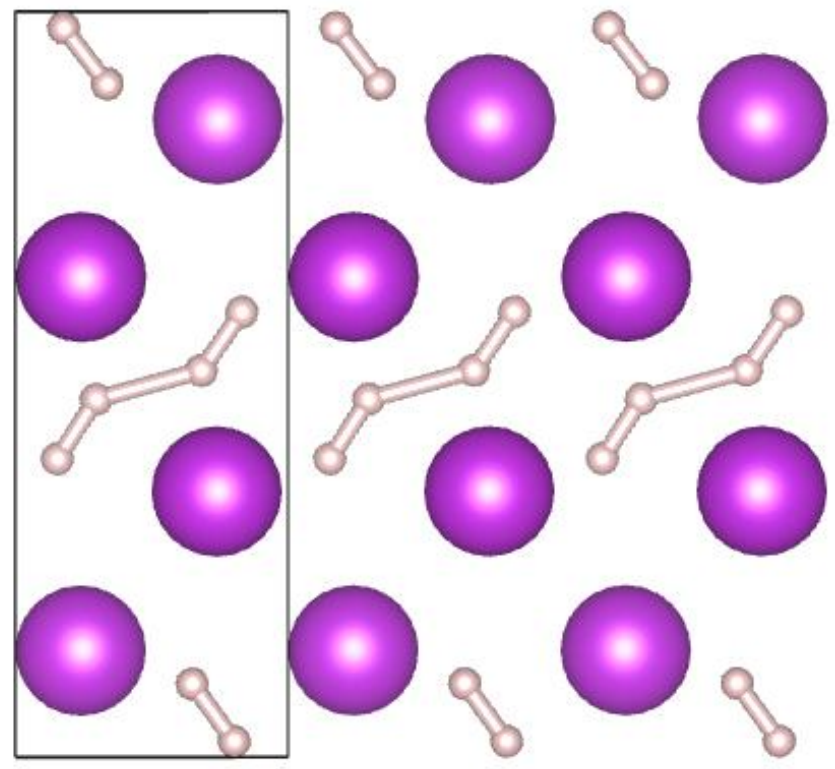

Space Group: P 1 (#1-1)
a = 3.12045 Å  α = 89.9511°
b = 8.32234 Å  β = 89.9437°
c = 3.04406 Å  γ = 89.9615°
V = 79.0524 $Å^3$

```
data_findsym-output

_symmetry_space_group_name_H-M 'P 21/n 21/n 2/m'
_symmetry_Int_Tables_number 58

_cell_length_a      3.04400
_cell_length_b      8.32200
_cell_length_c      3.12000
_cell_angle_alpha   90.00000
_cell_angle_beta    90.00000
_cell_angle_gamma   90.00000

loop_
_space_group_symop_operation_xyz
x,y,z
x+1/2,-y+1/2,-z+1/2
-x+1/2,y+1/2,-z+1/2
-x,-y,z
-x,-y,-z
-x+1/2,y+1/2,z+1/2
x+1/2,-y+1/2,z+1/2
x,y,-z

loop_
_atom_site_label
```

| | |
|---|---|
| | _atom_site_type_symbol<br>_atom_site_fract_x<br>_atom_site_fract_y<br>_atom_site_fract_z<br>_atom_site_occupancy<br>Bi1 Bi -0.24852 0.14380 0.00000 1.00000<br>H1 H -0.31012 0.47965 0.00000 1.00000<br>H2 H 0.15886 -0.40200 0.00000 1.00000 |
| *P1 (pseudo $P2_12_12_1$)-$BiH_2$* (155 GPa)<br>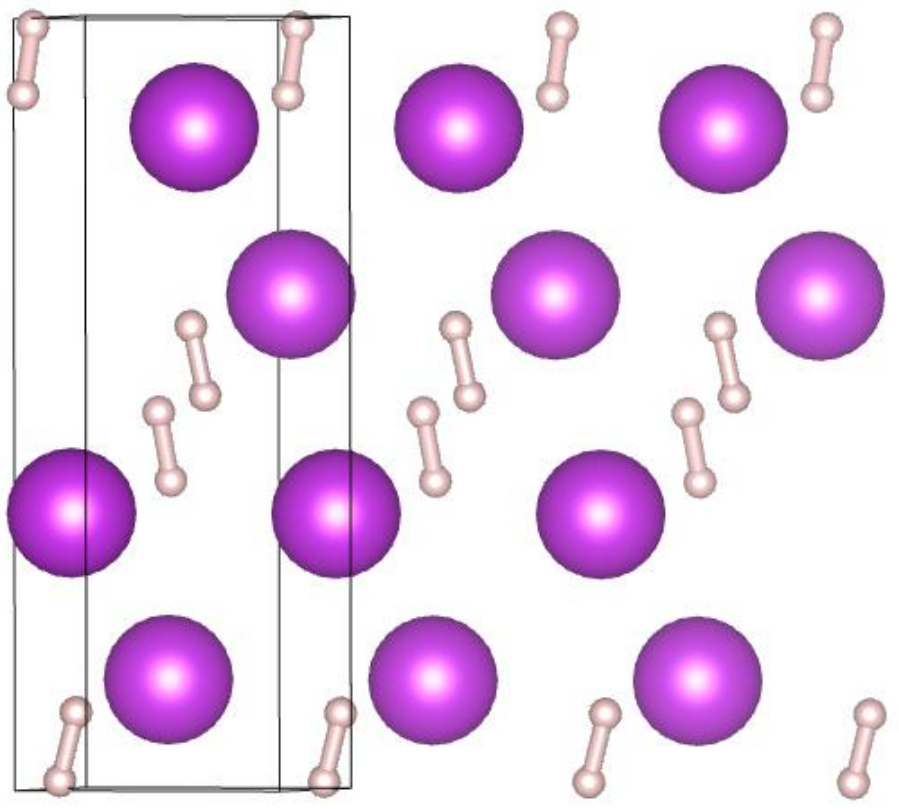<br>*$BiH_2$ molecular*<br>Space Group: P 1 (#1-1)<br>a = 3.08318 Å α = 89.9938°<br>b = 8.65121 Å β = 90.0419°<br>c = 2.96060 Å γ = 90.0807°<br>V = 78.9687 $Å^3$ | # CIF file<br># This file was generated by FINDSYM<br># Harold T. Stokes, Branton J. Campbell, Dorian M. Hatch<br># Brigham Young University, Provo, Utah, USA<br><br>data_findsym-output<br><br>_symmetry_space_group_name_H-M 'P 21 21 21'<br>_symmetry_Int_Tables_number 19<br><br>_cell_length_a 2.96100<br>_cell_length_b 8.65100<br>_cell_length_c 3.08300<br>_cell_angle_alpha 90.00000<br>_cell_angle_beta 90.00000<br>_cell_angle_gamma 90.00000<br><br>loop_<br>_space_group_symop_operation_xyz<br>x,y,z<br>x+1/2,-y+1/2,-z<br>-x,y+1/2,-z+1/2<br>-x+1/2,-y,z+1/2<br><br>loop_<br>_atom_site_label<br>_atom_site_type_symbol<br>_atom_site_fract_x<br>_atom_site_fract_y<br>_atom_site_fract_z<br>_atom_site_occupancy<br>Bi1 Bi -0.49893 -0.39330 0.01735 1.00000<br>H1 H 0.49373 0.34862 0.03091 1.00000<br>H2 H 0.47647 -0.24002 -0.47955 1.00000 |
| *$P2_1/m$-$BiH_2$* (150 GPa) [34] | _audit_creation_date 2024-10-14<br>_audit_creation_method 'Materials Studio'<br>_symmetry_space_group_name_H-M 'P21/M'<br>_symmetry_Int_Tables_number 11<br>_symmetry_cell_setting monoclinic<br>loop_<br>_symmetry_equiv_pos_as_xyz<br>x,y,z<br>-x,y+1/2,-z<br>-x,-y,-z<br>x,-y+1/2,z<br>_cell_length_a 4.5029<br>_cell_length_b 3.0751 |
| 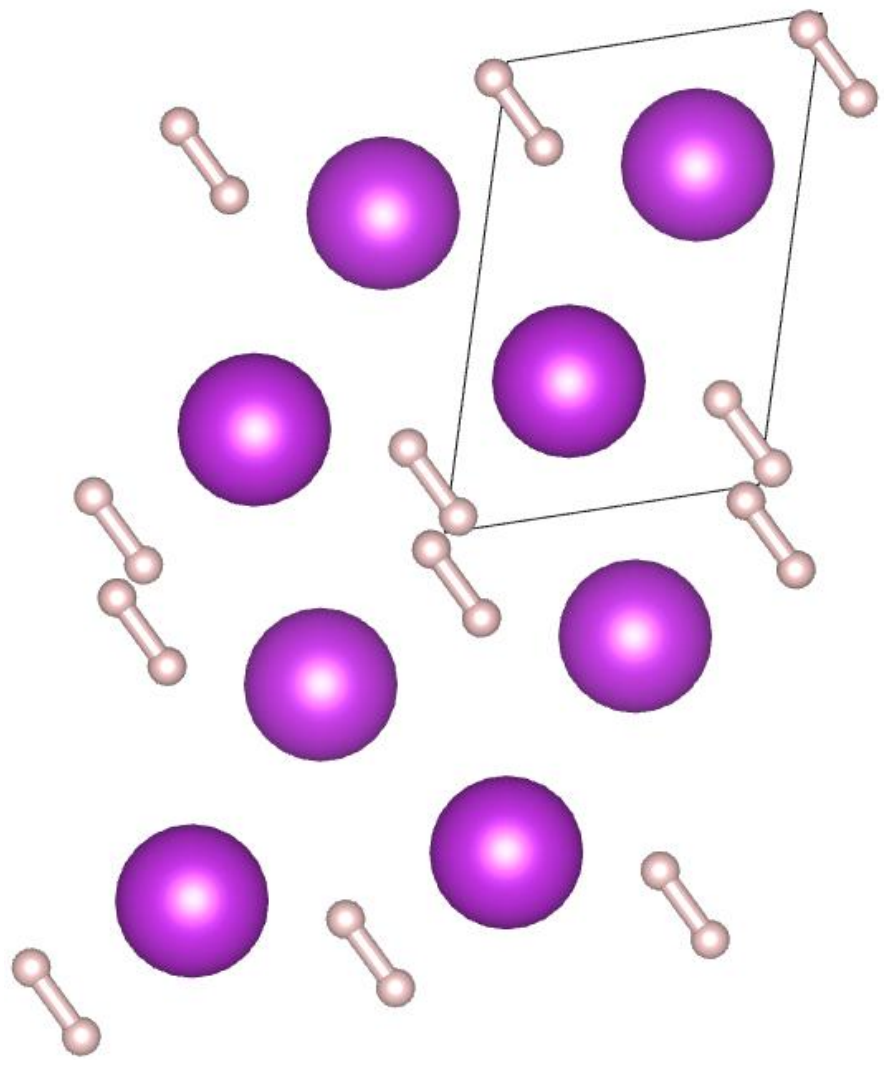<br>Space Group: P 21/m (#11-1)<br>a = 4.5029 Å α = 90.0°<br>b = 3.0751 Å β = 73.6949°<br>c = 3.0145 Å γ = 90.0°<br>V = 40.0626 Å$^3$ | _cell_length_c 3.0145<br>_cell_angle_alpha 90.0000<br>_cell_angle_beta 73.6949<br>_cell_angle_gamma 90.0000<br>_cell_volume 40.0626<br>loop_<br>_atom_site_label<br>_atom_site_type_symbol<br>_atom_site_fract_x<br>_atom_site_fract_y<br>_atom_site_fract_z<br>_atom_site_U_iso_or_equiv<br>_atom_site_adp_type<br>_atom_site_occupancy<br>Bi Bi 0.28727 0.25000 0.33760 0.00000 Uiso 1.00<br>H1 H 0.80375 0.25000 0.15578 0.00000 Uiso 1.00<br>H2 H 0.03102 0.75000 0.03594 0.00000 Uiso 1.00 |

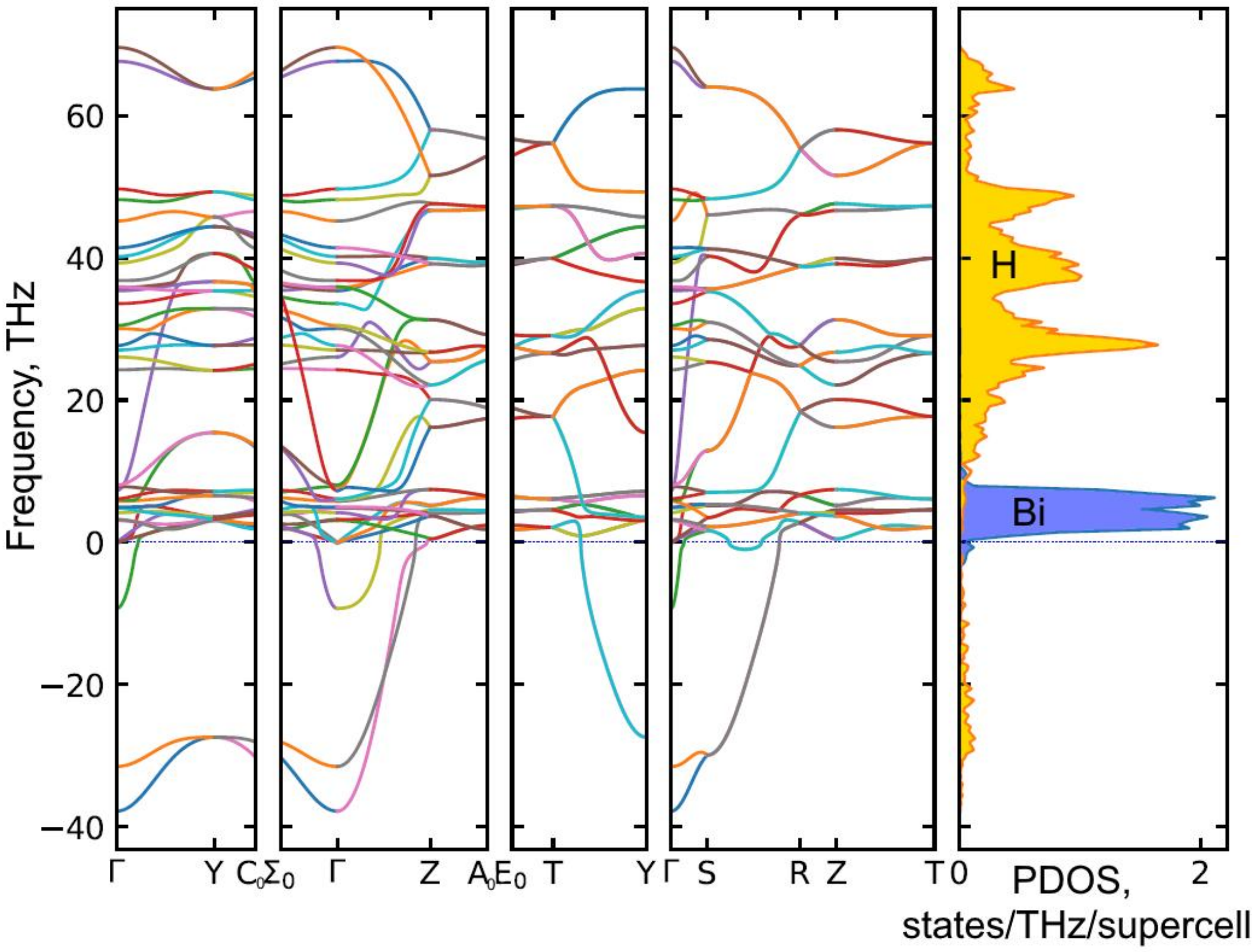

**Figure S13.** Phonon band structure and phonon density of states (PDOS) of *Cmcm*-$BiH_2$ (supercell $Bi_4H_8$) at 155 GPa. Harmonic approximation, VASP code.

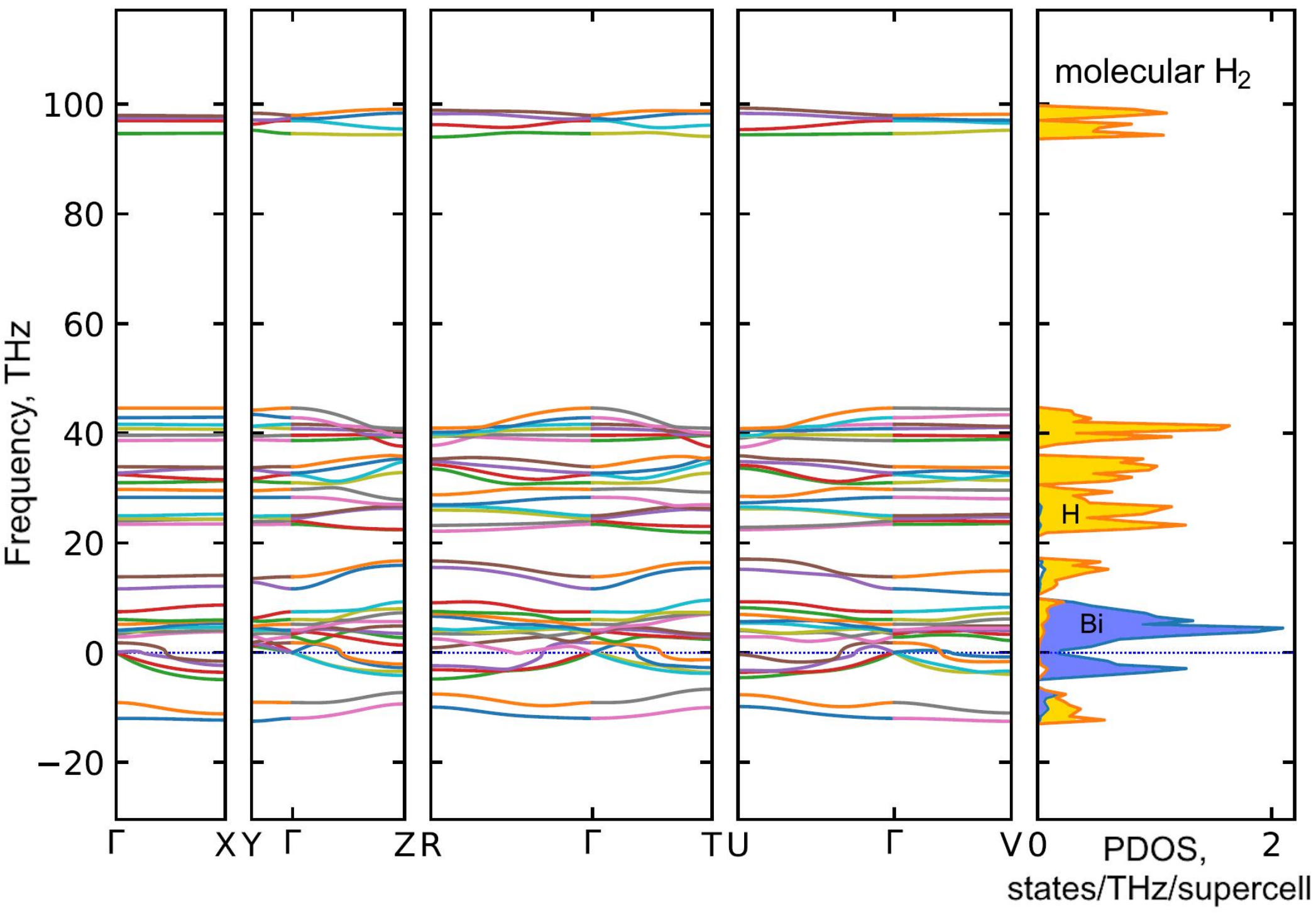

**Figure S14.** Phonon band structure and phonon density of states (PDOS) for *P*1 (pseudo *Pnnm*)-$BiH_2$ (supercell $Bi_4H_8$) at 155 GPa. Symmetrization with tolerance 0.2 transforms this structure to *Pnnm*. Harmonic approximation, VASP code.

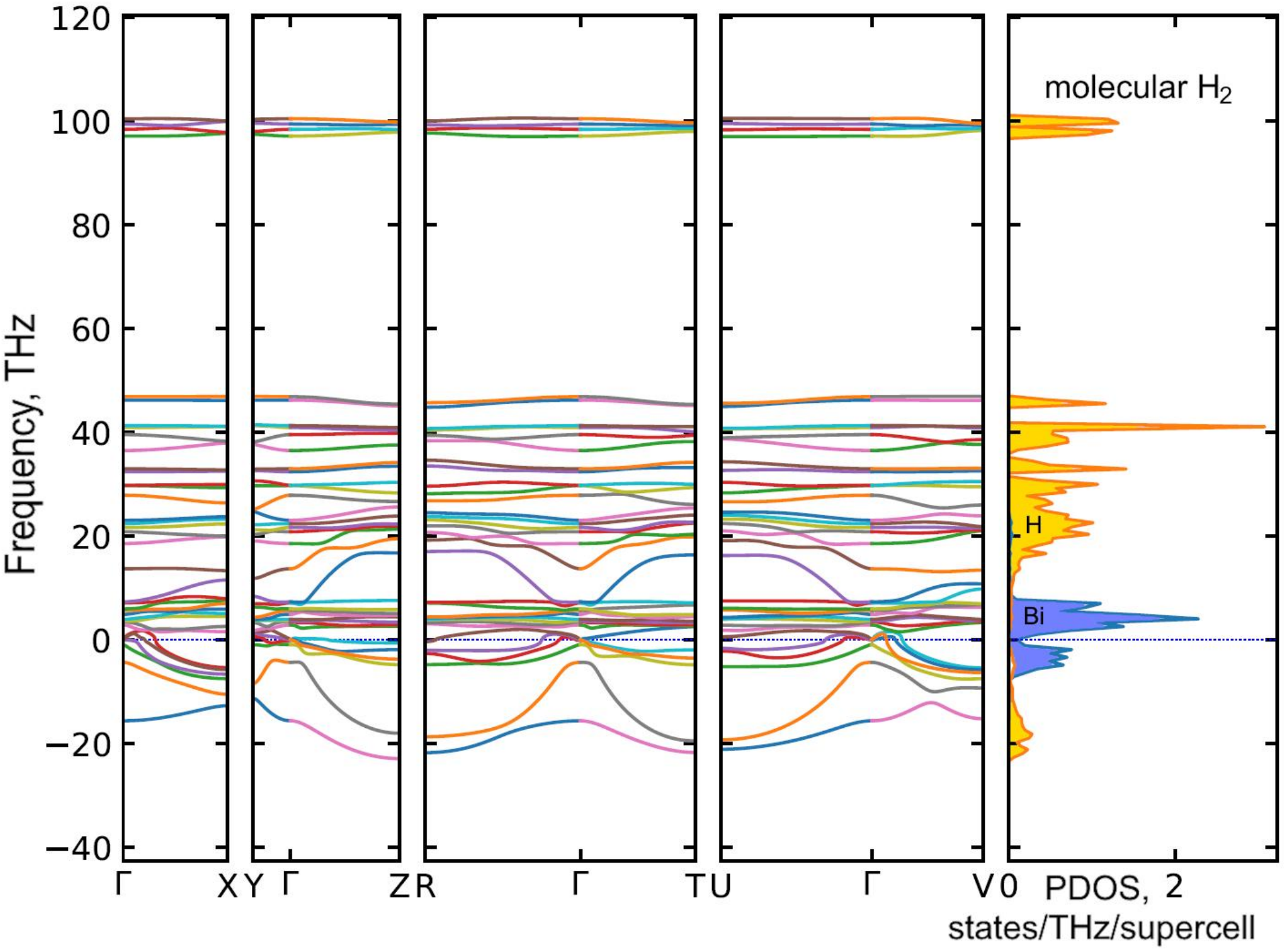

**Figure S15.** Phonon band structure and phonon DOS of completely molecular *P*1 (pseudo $P2_12_12_1$)-$BiH_2$ (supercell $Bi_4H_8$) at 155 GPa. Symmetrization with tolerance 0.2 transforms this structure to $P2_12_12_1$. Harmonic approximation, VASP code.

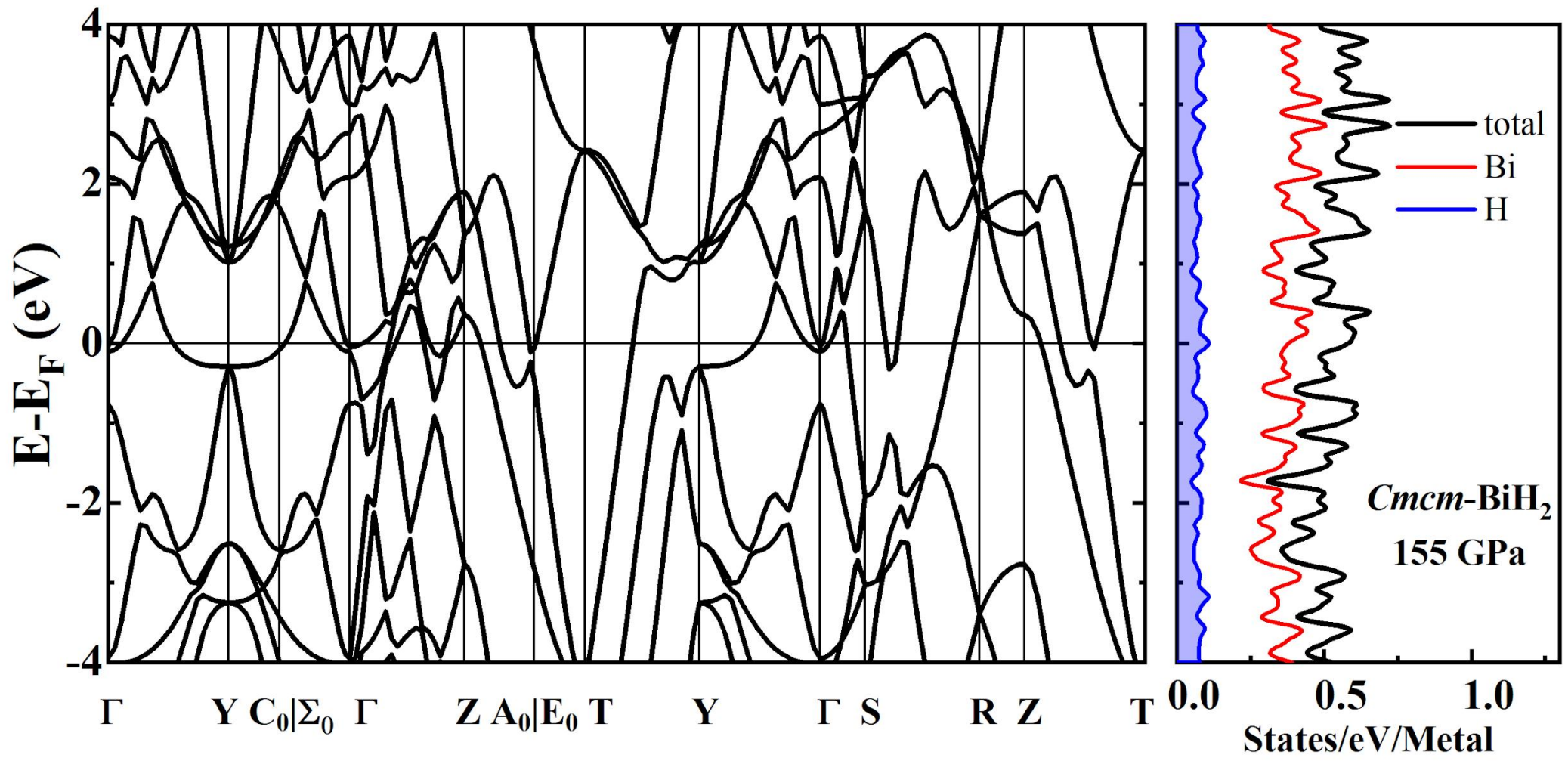

**Figure S16.** Electron band structure and density of states (DOS) of *Cmcm*-$BiH_2$ at 155 GPa.

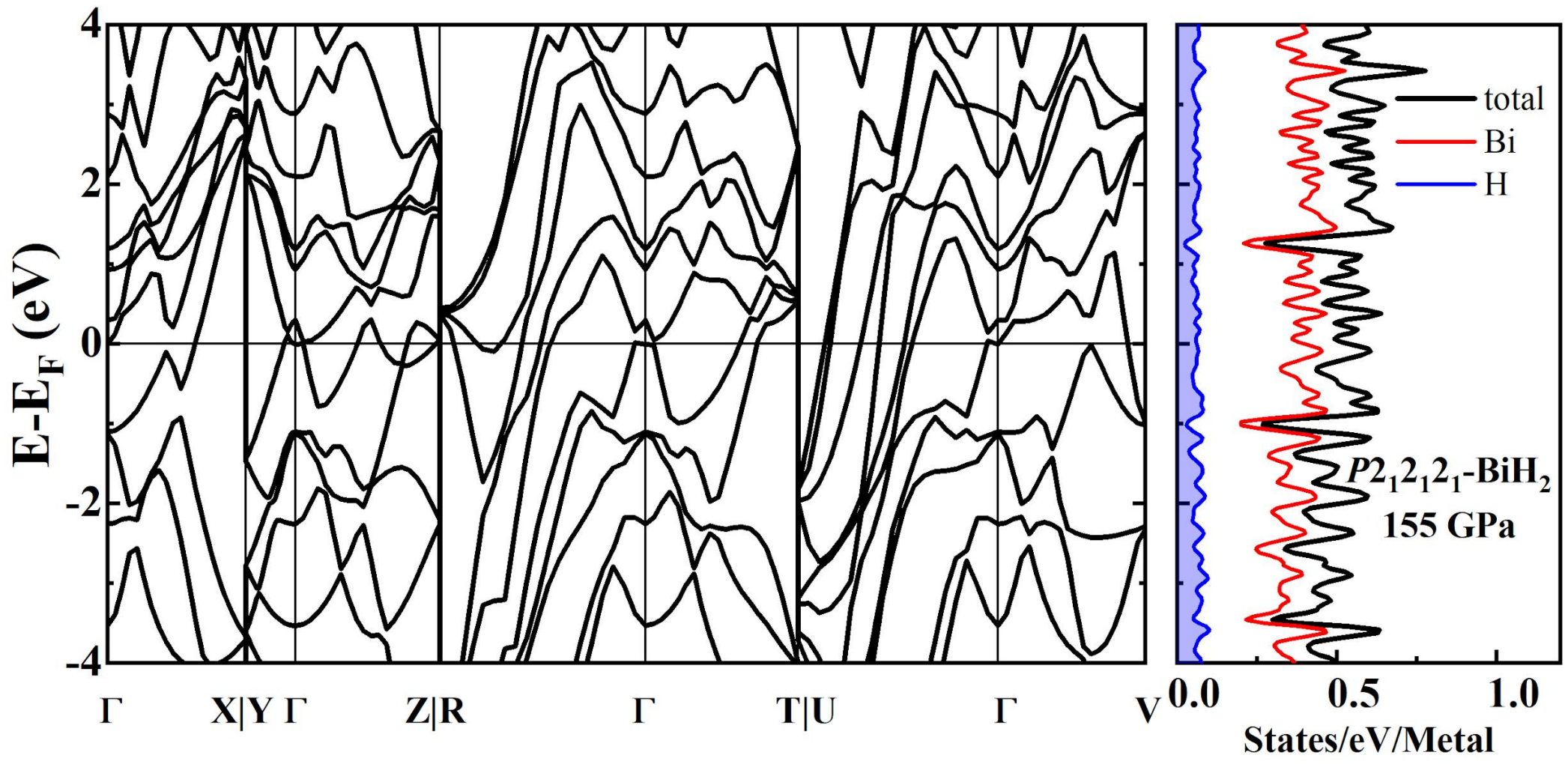

**Figure S17.** Electron band structure and density of states (DOS) of completely molecular *P*1 (pseudo *P*$2_12_12_1$)-$BiH_2$ at 155 GPa.

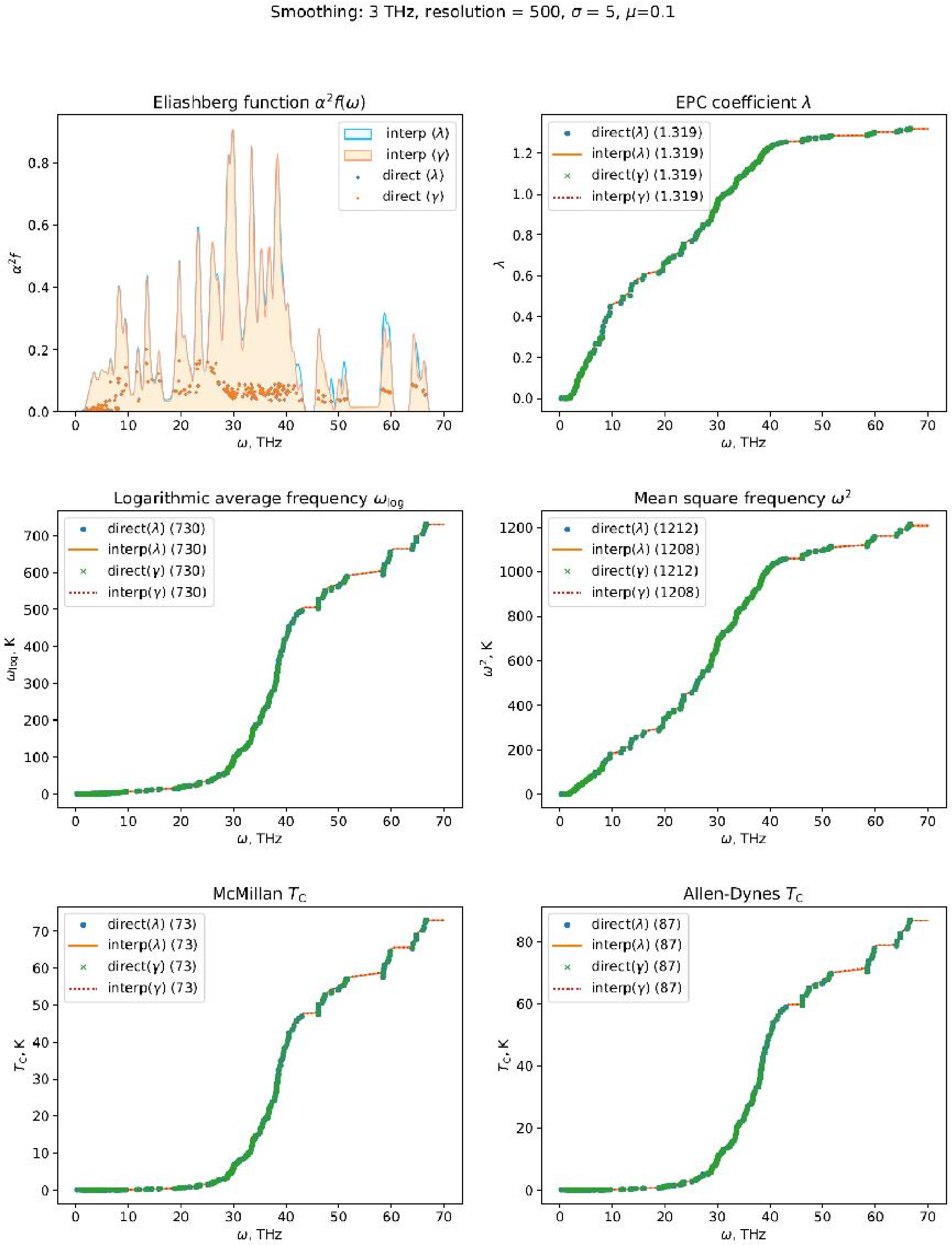

**Figure S18.** Frequency dependences of the electron-phonon interaction integrals for *Cmcm*-$BiH_2$ at 155 GPa ($\mu^*$ = 0.1). For calculations, we used a Python script written by G. Shutov [35,36].

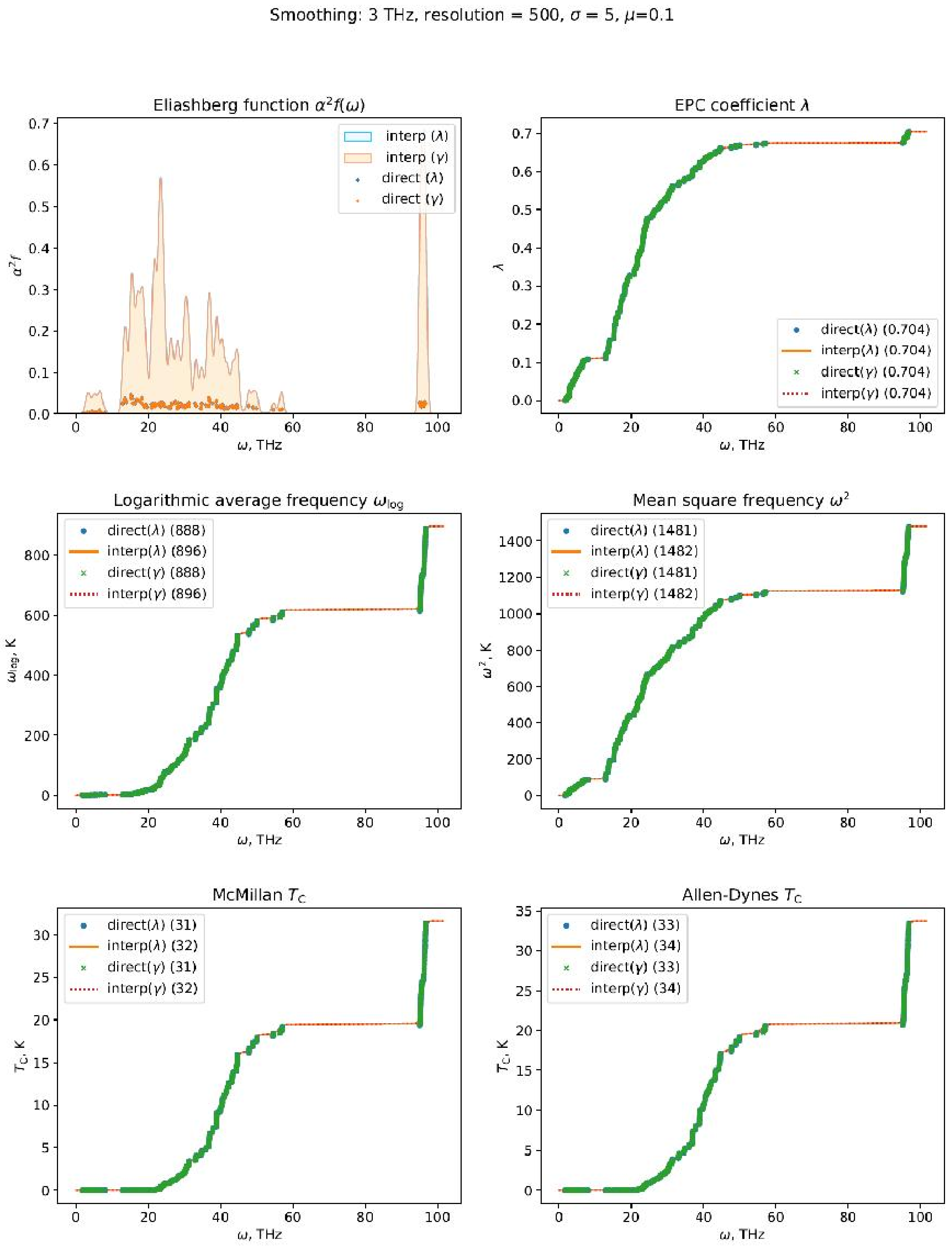

**Figure S19.** Frequency dependences of the electron-phonon interaction integrals for molecular $P1$ (pseudo $P2_12_12_1$)-$BiH_2$ at 155 GPa ($\mu^* = 0.1$). For calculations, we used a Python script written by G. Shutov [35,36].

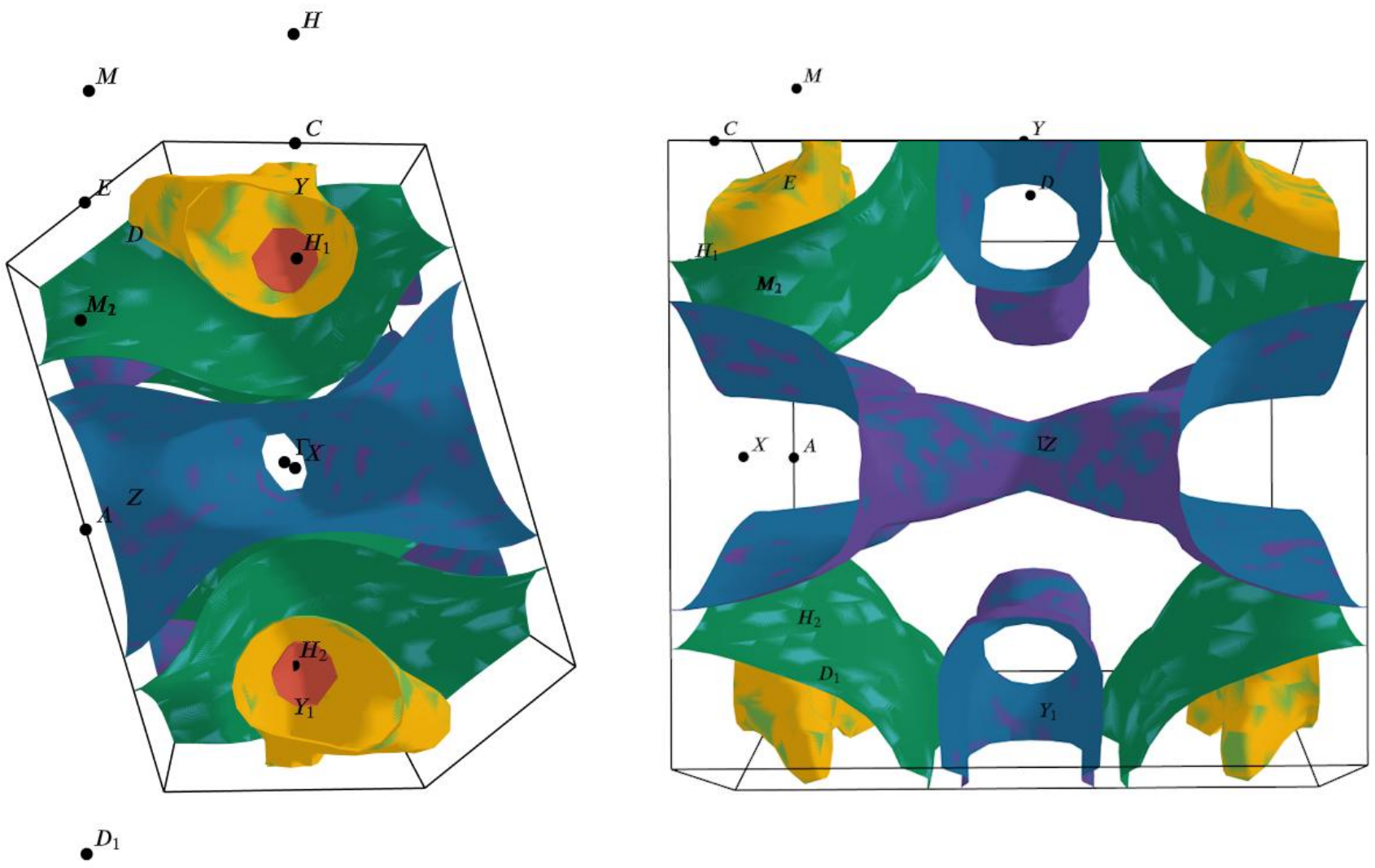

**Figure S20.** Profile and front view of the Fermi surface of $P2_1/m$-$BiH_2$ at 155 GPa (VASP, IFERMI [37]). Calculations were made with accounting of the spin-orbit coupling (SOC).

**Table S4.** Summary of the Fermi surface of $P2_1$/m-$BiH_2$ at 155 GPa. Calculations were made with accounting of the spin-orbit coupling (SOC).

Number of surfaces: 30
Area: 39.04 Å$^{-2}$
Avg velocity: 1.098e+06 m/s

| Band No. | Area, Å$^{-2}$ | Velocity avg, m/s | Dimensionality |
|---|---|---|---|
| 13 | 5.417 | 955600 | quasi-2D |
| 13 | 1.934 | 1060000 | 2 |
| 13 | 1.934 | 1060000 | 2 |
| 14 | 5.417 | 955600 | quasi-2D |
| 14 | 1.934 | 1060000 | 2 |
| 14 | 1.934 | 1060000 | 2 |
| 15 | 1.597 | 1408000 | 2 |
| 15 | 1.597 | 1408000 | 2 |
| 15 | 1.597 | 1408000 | 2 |
| 15 | 1.597 | 1408000 | 2 |
| 16 | 1.597 | 1408000 | 2 |
| 16 | 1.597 | 1408000 | 2 |
| 16 | 1.597 | 1408000 | 2 |
| 16 | 1.597 | 1408000 | 2 |
| 17 | 0.8801 | 851200 | 3 |
| 17 | 8.80E-01 | 851400 | 3 |
| 17 | 0.8801 | 851600 | 3 |
| 17 | 0.8801 | 851100 | 3 |
| 18 | 8.80E-01 | 851200 | 3 |
| 18 | 8.80E-01 | 851300 | 3 |
| 18 | 0.8801 | 851600 | 3 |
| 18 | 0.8801 | 851100 | 3 |
| 19 | 0.08127 | 530400 | 3 |
| 19 | 8.13E-02 | 530900 | 3 |
| 19 | 8.13E-02 | 530900 | 3 |
| 19 | 0.08127 | 530600 | 3 |
| 20 | 0.08127 | 530400 | 3 |
| 20 | 8.13E-02 | 530900 | 3 |
| 20 | 8.13E-02 | 530900 | 3 |
| 20 | 0.08127 | 530600 | 3 |

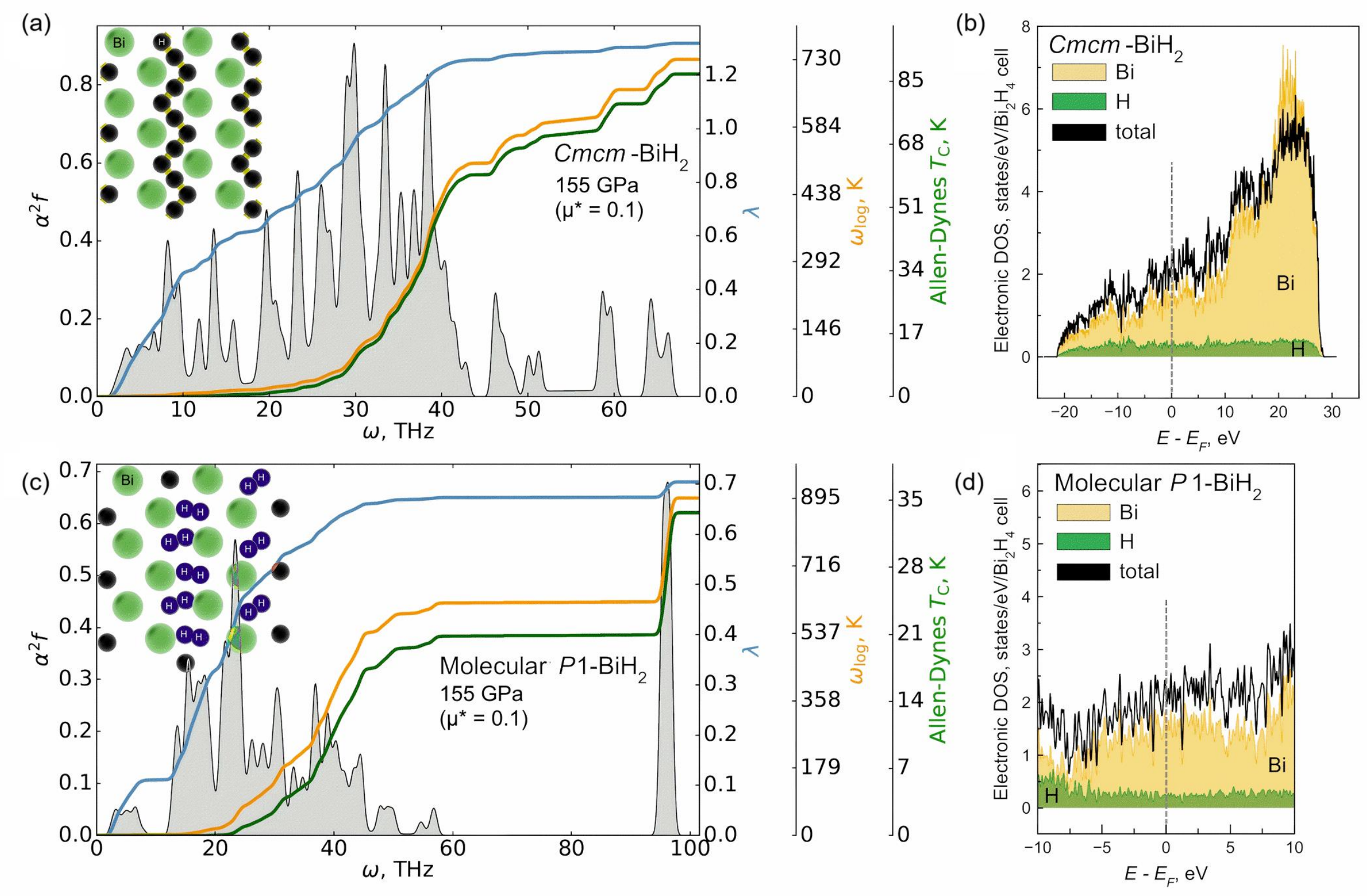

**Figure S21.** Calculations of superconducting properties and electron density of states in BiH2. Eliashberg function of (a) Cmcm-BiH2, and (c) P1-BiH2 at 155 GPa. The green curve corresponds to the critical temperature calculated by the Allen-Dynes formula with the Coulomb pseudopotential μ* = 0.1, the orange one – to the logarithmically averaged phonon frequency ωlog, and the blue one – to the electron-phonon interaction parameter (λ). Insets: crystal structures of Cmcm and P1-BiH2. Electron densities of states in (b) Cmcm-BiH2, and (d) molecular P1-BiH2 at 155 GPa.

**Table S5.** Summary of the Fermi surface of cubic $Fd\underline{3}m$-$Sn_8H_{30}$ at 200 GPa. Calculations were made with accounting of the spin-orbit coupling (SOC). Dimensionality is 3D for all bands.

| Number of surfaces: 132<br>Area: 19.07 Å$^{-2}$<br>Avg velocity: 1.038e+06 m/s | | |
|---|---|---|
| Band No. | Area, Å$^{-2}$ | Velocity avg, m/s |
| 70 | 0.07624 | 944100 |
| 70 | 0.07525 | 947600 |
| 70 | 0.07641 | 952000 |
| 70 | 0.07619 | 952800 |
| 70 | 0.0754 | 947900 |
| 70 | 0.07624 | 944100 |
| 71 | 0.1917 | 968500 |
| 71 | 0.1874 | 972200 |
| 71 | 0.3742 | 986700 |
| 71 | 0.4236 | 1076000 |
| 71 | 0.3742 | 986400 |
| 71 | 0.6232 | 1031000 |
| 71 | 0.1267 | 428500 |
| 71 | 0.09479 | 454200 |
| 71 | 0.2224 | 1238000 |
| 71 | 1.01E-04 | 127800 |
| 71 | 0.1855 | 960400 |
| 71 | 0.1874 | 972500 |
| 71 | 3.64E-04 | 224700 |
| 71 | 5.80E-04 | 431700 |
| 71 | 0.007809 | 302100 |
| 71 | 0.4117 | 1093000 |
| 71 | 0.001432 | 231900 |
| 71 | 1.77E-04 | 429500 |
| 71 | 1.01E-04 | 127800 |
| 71 | 0.001709 | 156000 |
| 71 | 0.224 | 1214000 |
| 71 | 8.72E-04 | 231800 |
| 71 | 3.70E-04 | 224200 |
| 71 | 2.332 | 1011000 |
| 71 | 0.007818 | 301800 |
| 71 | 0.001439 | 231700 |
| 71 | 1.78E-04 | 429700 |
| 71 | 5.86E-04 | 431500 |
| 71 | 0.2222 | 1239000 |
| 71 | 0.2226 | 1239000 |
| 71 | 0.09961 | 444700 |
| 71 | 0.003615 | 214500 |
| 71 | 0.00359 | 213800 |
| 71 | 0.003768 | 219700 |
| 71 | 0.09964 | 444800 |
| 71 | 0.001798 | 129400 |
| 71 | 0.4121 | 1092000 |
| 71 | 0.6238 | 1030000 |
| 71 | 0.3747 | 985900 |
| 71 | 0.001798 | 129400 |
| 71 | 0.1874 | 972700 |
| 71 | 0.1854 | 959700 |
| 71 | 0.4236 | 1076000 |
| 71 | 0.00377 | 217400 |
| 71 | 0.00359 | 213800 |
| 71 | 0.003616 | 212800 |
| 71 | 0.2239 | 1214000 |
| 71 | 0.3744 | 986400 |
| 71 | 5.86E-04 | 431500 |
| 71 | 0.007809 | 305200 |
| 71 | 1.78E-04 | 429700 |
| 71 | 0.1917 | 968500 |
| 71 | 0.001439 | 231700 |
| 71 | 0.1282 | 426400 |
| 71 | 3.70E-04 | 224200 |
| 71 | 0.2222 | 1238000 |
| 71 | 8.72E-04 | 231800 |
| 71 | 0.09478 | 454800 |
| 71 | 0.001709 | 156200 |
| 71 | 1.01E-04 | 127800 |
| 71 | 0.007799 | 305500 |
| 71 | 1.77E-04 | 429500 |
| 71 | 0.001432 | 231900 |
| 71 | 5.80E-04 | 431700 |
| 71 | 3.64E-04 | 224700 |
| 71 | 0.1874 | 972200 |
| 71 | 1.01E-04 | 127800 |
| 72 | 0.03291 | 880900 |
| 72 | 0.03181 | 871400 |
| 72 | 0.5922 | 1142000 |
| 72 | 0.5894 | 1131000 |
| 72 | 0.032 | 941300 |
| 72 | 0.03119 | 950300 |
| 72 | 0.03268 | 945200 |
| 72 | 0.03114 | 950300 |
| 72 | 0.03347 | 884900 |
| 72 | 0.03281 | 877700 |
| 72 | 0.03344 | 884300 |
| 72 | 0.03362 | 855500 |
| 72 | 0.03188 | 871300 |
| 72 | 0.03437 | 851200 |
| 72 | 0.5921 | 1143000 |
| 72 | 0.5891 | 1131000 |
| 72 | 0.5898 | 1131000 |
| 72 | 0.5921 | 1142000 |
| 72 | 0.4226 | 1001000 |
| 72 | 0.03341 | 880300 |
| 72 | 0.03425 | 851400 |
| 72 | 0.0319 | 870100 |
| 72 | 0.033 | 879300 |
| 72 | 0.03344 | 880900 |
| 72 | 0.03389 | 854200 |
| 72 | 0.59 | 1131000 |
| 72 | 0.592 | 1142000 |
| 72 | 0.03114 | 951100 |
| 72 | 0.03268 | 945200 |
| 72 | 0.03118 | 951100 |
| 72 | 0.032 | 941300 |
| 72 | 0.03291 | 876700 |
| 72 | 0.03203 | 870100 |
| 73 | 0.2107 | 1091000 |
| 73 | 0.2092 | 1091000 |
| 73 | 0.2106 | 1094000 |
| 73 | 0.2089 | 1091000 |
| 73 | 0.2089 | 1089000 |
| 73 | 0.2107 | 1093000 |
| 73 | 7.66E-04 | 45550 |
| 73 | 7.66E-04 | 45550 |
| 73 | 0.209 | 1090000 |
| 73 | 0.2107 | 1092000 |
| 74 | 0.07156 | 1024000 |
| 74 | 0.06914 | 1038000 |
| 74 | 0.07172 | 1025000 |
| 74 | 0.06895 | 1037000 |
| 74 | 0.06893 | 1038000 |
| 74 | 0.07174 | 1023000 |
| 74 | 0.06913 | 1040000 |
| 74 | 0.07152 | 1025000 |
| 75 | 0.01658 | 800400.0 |
| 75 | 0.01723 | 888800.0 |
| 75 | 0.01661 | 806100.0 |
| 75 | 0.01722 | 889000.0 |
| 75 | 0.01722 | 884200.0 |
| 75 | 0.01694 | 783700.0 |
| 75 | 0.01723 | 885100.0 |
| 75 | 0.01689 | 792100.0 |

76 (2015).